# Triplet Excitons and Associated Efficiency-Limiting Pathways in Organic Solar Cell Blends Based on (Non-) Halogenated PBDB-T and Y-Series

Jeannine Grüne[1,2]*, Giacomo Londi[3], Alexander J. Gillett[2], Basil Stähly[1], Sebastian Lulei[1], Maria Kotova[1], Yoann Olivier[3], Vladimir Dyakonov[1], and Andreas Sperlich[1]*

[1]Experimental Physics 6, Julius Maximilian University of Würzburg, Am Hubland, 97074 Würzburg, Germany

[2]Cavendish Laboratory, University of Cambridge, JJ Thomson Avenue, Cambridge, UK

[3]Laboratory for Computational Modeling of Functional Materials, Namur Institute of Structured Matter, University of Namur, Rue de Bruxelles, 61, 5000 Namur, Belgium

* jg2082@cam.ac.uk, sperlich@physik.uni-wuerzburg.de

**Abstract**

The great progress in organic photovoltaics (OPV) over the past few years has been largely achieved by the development of non-fullerene acceptors (NFAs), with power conversion efficiencies now approaching 20%. To further improve device performance, loss mechanisms must be identified and minimized. Triplet states are known to adversely affect device performance, since they can form energetically trapped excitons on low-lying states that are responsible for non-radiative losses or even device degradation. Halogenation of OPV materials has long been employed to tailor energy levels and to enhance open circuit voltage. Yet, the influence on recombination to triplet excitons has been largely unexplored. Using the complementary spin-sensitive methods of photoluminescence detected magnetic resonance (PLDMR) and transient electron paramagnetic resonance (trEPR) corroborated by transient absorption and quantum-chemical calculations, we unravel exciton pathways in OPV blends employing the polymer donors PBDB-T, PM6 and PM7 together with NFAs Y6 and Y7. All blends reveal triplet excitons on the NFA populated via non-geminate hole back transfer and, in blends with halogenated donors, also by spin-orbit coupling driven intersystem crossing. Identifying these triplet formation pathways in all tested solar cell absorber films highlights the untapped potential for improved charge generation to further increase plateauing OPV efficiencies.

## 1. Introduction

Organic solar cells (OSC) based on non-fullerene acceptors (NFAs) have attracted much attention due to their strong absorption in the visible and near infrared spectral regions and their energy-level tunability, in contrast to fullerene acceptors.[1] However, despite the great success of OSC development to power conversion efficiencies (PCE) up to 19%[2], there is still untapped potential for further improvement. One limiting factor is the total voltage loss ($\Delta V_{loss} = q^{-1}E_g - V_{OC}$), defined as the difference between the optical band gap ($E_g$) and the open-circuit voltage ($V_{OC}$).[3-6] Besides, $\Delta V_{loss}$ is composed of losses due to radiative ($\Delta V_r$) and non-radiative ($\Delta V_{nr}$) recombination.[4, 7] According to the Shockley-Queisser limit, an ideal solar cell possesses only radiative recombination, making the contribution $\Delta V_r$ unavoidable.[8, 9] Non-radiative recombination processes thus reduce $V_{OC}$ and the short-circuit current and are therefore of great interest for OSC research.[3, 5] Furthermore, after optical excitation of donor (D) or acceptor (A) moieties, charge transfer (CT) at the interface is expected to reduce the photovoltage due to the difference in the D or A band gaps and the energy of CT states ($E_g - E_{CT}$).[6, 10, 11] However, in state-of-the-art D:A combinations, such as PM6:Y6, this difference is almost negligible, making non-radiative recombination the most dominant in terms of efficiency losses.[7]

Due to the close energetic alignment between NFA singlet states and the CT states, especially in the PBDB-T:Y-series blends, NFA triplet states are energetically located below the CT states and represent a significant non-radiative decay channel for excitons.[12] Preventing the population of these triplet states is extremely difficult, even in state-of-the art blends as shown in this work. Direct intersystem crossing (ISC) from optically-excited donor or acceptor singlet states to molecular triplet states is one of the possible triplet formation mechanisms.[13, 14] The yield of this geminate pathway depends on the number of optically-generated singlet excitons not reaching the D:A interface within their exciton diffusion length to undergo charge transfer.[15, 16] Another loss pathway is due to triplet CT states that relax spin-allowed to energetically lower molecular triplet states, also called electron or hole back transfer (EBT, HBT).[3, 17] In polymer- or fullerene-based blends, triplet states are thereby often formed not by ISC from the $S_1$ state but instead by geminate back transfer directly following the charge transfer.[18, 19] In newer materials, such as NFA-based blends, HBT from CT states formed after non-geminate recombination was shown to have a considerable impact by limiting $V_{OC}$.[3] All triplet formation mechanisms can lead to triplet excitons, which are localized on the lowest-lying molecular triplet states and thus increase non-radiative recombination.

In the present work, we set the focus on the impact of donor and acceptor halogenation on the formation of triplet excitons. Halogenation of, e.g. donor polymers can improve device efficiencies, as shown for fluorination and chlorination of PBDB-T to yield PM6 and PM7, respectively.[20-22] Thereby, halogenation is beneficial in terms of PCE, since it stabilizes the highest occupied molecular orbital (HOMO) energy level, resulting in increased $V_{OC}$.[20, 23] However, the question arises whether the reduced HOMO offset impacts the important hole transfer process in NFA blends. Another advantage of halogenation is observed for morphology since halogenation increases the molecular planarity and ordering. This in turn results in improved aggregation and larger domain sizes, leading to a rise in charge carrier mobility and in fill factor but could also lead to undissociated singlet excitons that undergo ISC.[24] Additionally, ISC rates of organic semiconductors could be enhanced substantially by halogenation as shown for organic light-emitting diodes (OLEDs) employing thermally activated delayed fluorescence (TADF), where bromination or iodination increases reverse ISC rates due to the heavy atom effect.[23] In this context, it remains to be clarified whether fluorination and chlorination of OPV materials also have an impact on ISC and hole transfer kinetics or overall on triplet exciton formation.

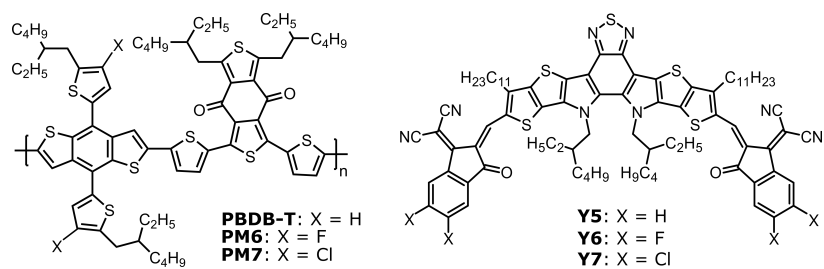

**Figure 1.** Chemical structures of studied organic solar cell materials. As donors PBDB-T (X = H), fluorinated version PM6 (X = F, also called PBDB-T-2F) and chlorinated version PM7 (X = Cl, also called PBDB-T-2Cl) are investigated. Acceptors represent Y5 (X = H, also called BTP), Y6 (X = F, also called BTP-4F) and chlorinated version Y7 (X = Cl, also called BTP-4Cl).

Since formation of long-lived triplet states results in a loss of OSC performance and can exacerbate device degradation, studying the presence of triplet excitons and their generation mechanisms is essential. We used two complementary methods, which are highly sensitive to triplet excitons: photoluminescence detected magnetic resonance (PLDMR) and transient electron paramagnetic resonance (trEPR). Both methods were often applied independently of each other in the past to investigate triplet formation in OSCs.[3, 25-28] Especially the combination of both techniques has been shown to be very powerful: trEPR probes triplet excitons directly by detection of microwave absorption between triplet sublevels occurring during resonance transitions and reveals highly spin-polarized, i.e., geminate triplet exciton formation pathways, such as ISC or geminate HBT.[29] In contrast, PLDMR probes triplet states indirectly with the higher sensitivity of optical detection and thus reveals those spin states that are associated with luminescence, e.g. via triplet-triplet annihilation (TTA), ground state depletion or (reverse) ISC.[30, 31] PLDMR is therefore well-suited to detect long-lived triplet excitons including those formed by non-geminate recombination and identify their molecular localization.[26, 27] To further study the influence of halogenation on excited state kinetics, including hole transfer and ISC rate constants, we performed transient absorption (TA) measurements and quantum-chemical (QC) calculations. By the combination of these complementary techniques, we thoroughly investigated the spin physics of state-of-the-art donor and acceptor materials in different combinations: the polymer PBDB-T and its halogenated (i.e., fluorinated and chlorinated) variants PM6 and PM7, as well as the respective NFAs Y6 and Y7 (Figure 1, whereby Y5 is used for comparison of (non-) halogenation in QC calculations).

## 2. Probing Triplet Excitons with Magnetic Resonance

Triplet excitons (either CT or molecular) are characterized by two interacting electron spins, whose fundamental spin physics is briefly described in the following. The corresponding Hamiltonian operator in X-band regime (microwave frequency $\nu_{MW} \approx 9.4$ GHz and external magnetic field $B_0 \approx 340$ mT) can be reduced to mainly three contributions: the electron Zeeman (EZ) interaction $\hat{H}_{EZ}$, zero-field splitting (ZFS) $\hat{H}_{ZFS}$ and exchange interaction $\hat{H}_{EX}$ (other contributions, including hyperfine fields, can be considered as negligible perturbations):[32]

$$\hat{H} = \hat{H}_{EZ} + \hat{H}_{ZFS} + \hat{H}_{EX} = g\,\mu_B\,\hat{\vec{S}}\vec{B} + \hat{\vec{S}}^T \boldsymbol{D}\, \hat{\vec{S}} + \hat{\vec{S}}_1^T \boldsymbol{J}\, \hat{\vec{S}}_2$$

Here, $\hat{\vec{S}}$ is the total spin angular momentum operator (with $\hat{\vec{S}}_1$ and $\hat{\vec{S}}_2$ being the operators for the respective electron spins), $\boldsymbol{g}$ the g-tensor (assumed to be isotropic due to a small spin-orbit coupling (SOC) in organic molecules) and $\mu_B$ the Bohr magneton, $\boldsymbol{D}$ the ZFS tensor and $\boldsymbol{J}$ the exchange-interaction tensor. The Zeeman interaction describes the quantized splitting of the paramagnetic states into their sublevels with spin quantum numbers $m_s$ = +1, 0, -1, i.e., the spin Hamiltonian eigenstates $|T_-\rangle$, $|T_0\rangle$ and $|T_+\rangle$, based on the interaction with the external magnetic field $\vec{B}$.[33] For nearby electron and hole (i.e., a locally excited state on a given molecule), the triplet sublevels are already energetically split in absence of an external magnetic field due to dipolar interactions between the spins of the unpaired electrons. This splitting is referred to as ZFS with the two scalar parameters $D$ and $E$ of the corresponding ZFS tensor $\boldsymbol{D}$: the axial parameter $D$ is related to the average inter-spin distance $r$ (to first order approximation $D \sim r^{-3}$), whereby $E$ describes the rhombicity, thus the deviation from axial symmetry.[32, 34, 35] For CT states with larger spin-spin distances than those of molecular states, spin conservation during charge separation generates spin-correlated radical pairs (SCRP) with exchange $J$ and dipolar interaction $D$ being in a comparable range. The Hamiltonian eigenstates are then represented by pure $|^3CT_-\rangle$ and $|^3CT_+\rangle$ triplet states ($m_s$ = ±1) and two mixed singlet-triplet states $|^3CT_0\rangle$ and $|^1CT_0\rangle$ ($m_s$ = 0), whereby due to angular momentum conservation, only a spin mixing between the latter two can occur. Further information about triplet spin states is given in SI.

## 2.1 Detecting Triplet Excitons with Spin-Sensitive Photoluminescence

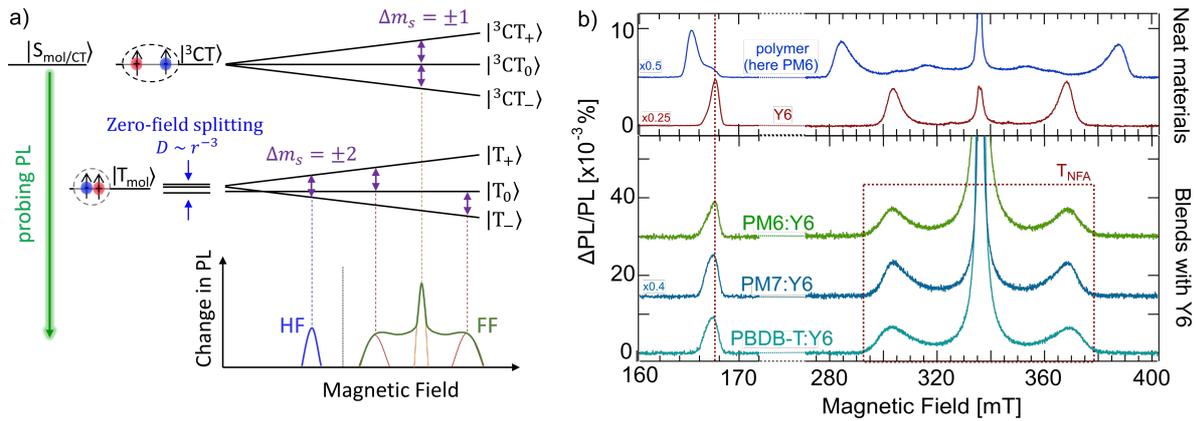

**Figure 2.** Photoluminescence detected magnetic resonance (PLDMR) of triplet excitons. a) PLDMR probes PL changes of singlet states ($|S_{mol/CT}\rangle$), which are coupled to paramagnetic triplet states. CT and molecular triplet states ($|^3CT\rangle$, $|T_{mol}\rangle$) are split due to Zeeman and ZFS interactions: $|^3CT_{+,0,-}\rangle$, $|T_{+,0,-}\rangle$. Microwave irradiation induces $\Delta m_s = \pm 1$ transitions (purple), leading to a change in the PL yield detected as full field spectrum (FF). An additional signal is detected at half the magnetic field (HF) for $\Delta m_s = \pm 2$ transitions. b) PLDMR spectra for neat materials PM6 and Y6 as well as the blends PBDB-T:Y6, PM6:Y6 and PM7:Y6. The spectral width of the FF signal and the position of the HF signal are determined by the material-dependent ZFS and therefore the triplet excitons in the blends can be assigned to the NFA Y6 (red dotted box). All PLDMR measurements used $\lambda_{ex}$ = 473 nm for excitation of donor and acceptor and were performed at $T$ = 10 K.

**Figure 2a** depicts the PLDMR principle of probing the PL change of singlet states (either CT or molecular), which are coupled (via TTA, ISC, etc.) with paramagnetic states.[28] Microwave irradiation induces transitions between the sublevels (purple arrows) leading to a change in PL intensity. In the case of $|^3CT\rangle$ states, the comparatively large inter-spin distance results in negligible dipolar interaction (negligible $D$ value), which in turn induces a symmetrical splitting of triplet sublevels $|^3CT_+\rangle$, $|^3CT_0\rangle$ and $|^3CT_-\rangle$ in a magnetic field. Microwave irradiation of frequency $\nu_{MW}$, resonant to the energetic sublevel splitting, induces spin-allowed $\Delta m_s = \pm 1$ transitions at closely spaced magnetic field values. The result is a change of steady-state triplet population, leading to an increased ($\Delta PL/PL > 0$) or reduced ($\Delta PL/PL < 0$) PL yield under resonance conditions, referred to as full-field (FF) signal. For the case of a localized molecular triplet ($|T_{mol}\rangle$ in Figure 2a), the ZFS results in an asymmetric splitting of the triplet sublevels in the applied magnetic field. Thus, microwave irradiation induces transitions and PL changes at more widely spaced magnetic field values, the distance of which depends linearly on the ZFS parameters $D$ and $E$. In addition, there is a certain probability of $\Delta m_s = \pm 2$ transitions at half of the resonant magnetic field, referred to as half-field (HF) signal.[34] The intensity of the HF signal increases quadratically with $D$ (intensity $\sim D^2 \sim r^{-6}$) and is therefore detectable for molecular triplet excitons only.[36] The ZFS parameter $D$, related to the total width of the spectrum ($|2D|/g\mu_B$), together with the spectral position of the HF signal allows for the molecular localization of the triplet states involved in the PL generation process to be readily determined.[27]

**Figure 2b** presents PLDMR spectra of the pristine materials PM6 and Y6 as well as of blends of Y6 with the polymer donors PBDB-T, PM6 and PM7. The measurements were performed for thin-film samples produced by an optimized spin-coating recipe comparable to OSC device production (recipe given in Experimental Section, measurements for drop-cast samples in EPR tubes are shown in Figure S11 and S12).[37, 38] The PLDMR spectra of the pristine polymers are almost identical; the spectrum of PM6 is shown here while the spectra of PBDB-T and PM7 can be found in Figure S9. All polymer spectra show a broad molecular triplet feature with $D/h$ = 1500 MHz (110 mT width) and a HF signal at $B$ = 165 mT. Pristine NFA Y6 shows a molecular triplet signal with a smaller $D$ value of $D/h$ = 950 MHz (70 mT width) and a HF signal at $B$ = 166.8 mT. Spectral simulation parameters derived by the MATLAB toolbox EasySpin[39] are summarized in Table S5. PLDMR spectra of all neat materials show additional central narrow peaks, which in polymers are generally assigned to photogenerated polaron pairs.[40-43] Neat NFAs, as Y6 or ITIC, show negligible photogeneration of polarons, as evidenced by light induced EPR (LEPR), suggesting short-lived (intermolecular) excited states with weak dipolar interactions, such as SCRPs, being the origin of the central peak in PLDMR.[42-44] When blending Y6 with the donors PM6, PM7 and PBDB-T (lower traces), the intensity of the central peak increases significantly due to the increased formation of separated polaron pairs and spectrally overlapping D:A interfacial CT states. Since PLDMR is performed on thin films without electrodes, all formed CT states and polaron pairs do eventually recombine. By means of trPLDMR, we can assign the PLDMR response of the NFA triplet and CT states to be positive (Figure S10), mostly stemming from annihilation processes such as TTA or triplet-polaron annihilation (TPA).[26, 45, 46] Indeed, TPA was found to be one of the dominant bimolecular interactions in PM6:Y6, responsible for significant voltage losses.[3, 41, 47]

All three blends show a broad triplet feature (300 – 370 mT) arising from long-lived molecular triplet excitons. The axial ZFS parameter $D$, i.e., the width of the FF spectra, and the positions of the HF signals at 166.8 mT (dotted vertical line) clearly assign these PLDMR features to triplet excitons on the NFA Y6. The $D$ value in the blends is slightly larger with $D/h$ = 1020 - 1040 MHz (Table S5) than in neat Y6 due to a marginally different molecular packing. However, the spin density delocalization is not strongly affected according to the relation $D \sim r^{-3}$. Furthermore, the triplet features in all PLDMR spectra can only be simulated taking considerable molecular ordering into account, leading to pronounced spectral "wings".[39, 48] The ordering factor $\lambda_\Theta$ gives information about the preferred molecular orientation relative to the external magnetic field, defined by the angle Θ. Neat Y6 shows a clear preferential alignment of the molecules (Table S5), in line with intermolecular face-on stacking and face-on stacking on the substrate, as evidenced by GIWAXS.[49-51] When blending Y6 with the polymers, the ordering is comparable for all blends (Table S5). The preferential orientation of Y6 molecules in the blends supports that the backbone ordering known for neat Y6 is maintained when blended with the polymer, as already confirmed for PM6:Y6 by GIWAXS measurements.[52] The PLDMR spectra suggest similar preferential orientation of Y6 also with PBDB-T and PM7, whereby this face-on orientation in the blend is shown to be beneficial for charge transport in the direction normal to the substrate surface.[53, 54]

The triplet excitons detected on the NFA can be generated either by direct ISC, non-geminate HBT or by geminate HBT. Although PLDMR is not able to distinguish between these formation mechanisms, its high sensitivity is crucial to reveal the presence of triplet excitons generated by all efficiency-limiting pathways. This sensitivity is achieved due to the optical detection and the continuous illumination, enhancing spin polarization due to annihilation effects.[29] PLDMR measurements were also performed with NFA Y7 (Figure S12), where triplet excitons were found on Y7. Hence, we observe molecular triplet excitons in all studied OPV blend, localized on the NFA.

## 2.2 Probing Triplet Pathways with Transient EPR

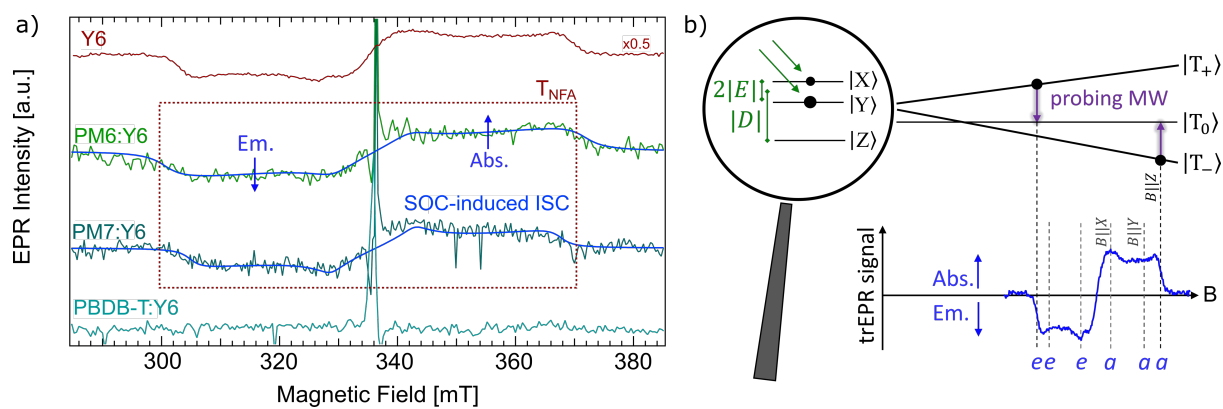

**Figure 3.** trEPR spectra of pristine Y6, PM6:Y6, PM7:Y6 and PBDB-T:Y6. a) trEPR spectra of blends with the halogenated donors (PM6 and PM7) have the same spectral fingerprint as triplets on pristine Y6 (red dotted box) as shown by the spectral fits (blue). The blend of PBDB-T:Y6 shows no detectable triplet formation. b) Formation of polarized trEPR spectra for spin-orbit coupling (SOC) driven ISC triplets. ISC is sublevel selective and acts on the zero-field sublevels $|X\rangle$, $|Y\rangle$ and $|Z\rangle$. These populations are converted into high-field populations $|T_+\rangle$, $|T_0\rangle$ and $|T_-\rangle$, depending on the three principal axes orientation of the ZFS tensor $\boldsymbol{D}$ with respect to magnetic field $\vec{B}$. Microwave emission (*e*) or absorption (*a*) for $\Delta m_s = \pm 1$ transitions (purple) results in the shown Y6 trEPR pattern (blue). The *eeeaaa* polarization pattern is due to SOC-driven ISC. All trEPR measurements used $\lambda_{ex}$ = 532 nm for excitation of donor and acceptor and were performed at *T* = 10 K.

While PLDMR probes steady-state populations under continuous illumination, trEPR probes short-lived spin polarization generated by pulsed laser excitation. Thereby, trEPR is able to reveal generation pathways of geminate triplet excitons with distinct spin polarization due to specific mechanisms, such as ISC.[33, 34] As discussed below, the fraction of these geminate triplet excitons is very low in the studied blends. For the spin-coated samples, this yields a signal that is detectable but too weak to determine the spin polarization pattern with confidence (Figure S2). Therefore, trEPR measurements were additionally performed on drop-cast samples (Figure 3), which yielded a better signal-to-noise ratio to reliably identify the population mechanism (see SI).

**Figure 3a** shows the trEPR spectrum of Y6 and its blends with the three donor polymers (trEPR of pristine polymer donor as well as blends with Y7 are shown in Figure S15 and Figure S16). Neat Y6 shows a 70 mT wide spectrum, corresponding to $D/h$ = 945 MHz, similar to the value obtained with PLDMR ($D/h$ = 950 MHz). When blending Y6 with the polymers, all blends show an additional intense middle signal with narrow microwave emission/absorption feature ($B$ = 336.5 mT), corresponding to interfacial CT states and photogenerated polarons. The width correlates with that of PLDMR middle features in Figure 2b, assigning both to the same $|^3CT\rangle$ states. Regarding the molecular (broad) trEPR feature, almost identical spectral fingerprints are detected for the blends of PM6:Y6 and PM7:Y6. The same spectral width proves the signals are also arising from Y6 triplet excitons as already shown with PLDMR. In contrast to PLDMR, trEPR spectra exhibit no visible molecular ordering, even in the spin-coated blends.[29] This is attributable to the fact that trEPR probes all (highly) spin-polarized triplet excitons, while PLDMR probes predominantly those triplets that are associated with luminescence, e.g. via more dominant TTA in crystalline phases as a result of an increased diffusion length.[29, 55, 56] The polarization pattern of the trEPR signal of PM6:Y6 and PM7:Y6 displays an *eeeaaa* signature, i.e., microwave emission at lower and microwave absorption at higher fields, consistent with triplet excitons populated by the geminate pathway of SOC-driven ISC.

Regarding PBDB-T:Y6, molecular triplet excitons on Y6 are detected with PLDMR (Figure 2b), whereby no molecular triplet feature in trEPR is observed. While trEPR probes geminate triplet excitons only, PLDMR is also sensitive to triplet excitons formed by non-geminate recombination. This finding assigns the triplet excitons in PBDB-T:Y6 visible in PLDMR (Figure 2) to non-geminate HBT. As also discussed below, non-geminate triplet excitons represent the main contribution of energetically trapped triplet excitons. Thus, the triplet yield visible in trEPR is overall a very minor triplet channel but important to proof the mechanism of present SOC-driven ISC.

**Figure 3b** shows the formation of the trEPR spectral pattern of the spin-polarized triplet feature of the neat materials and the blends. It arises by the selective population of the triplet sublevels by ISC from the excited singlet states.[34] This ISC mechanism for molecular excitons with small wave function extent (large $D$ value) is based on SOC, which is dominant for small electron-hole distances.[57] At larger inter-spin distances, SOC is negligible and (reverse) ISC is driven by hyperfine interactions (HFI) leading to different trEPR spectral patterns for, e.g. triplet excitons formed via geminate HBT, discussed further in the SI.[3, 58] SOC-driven ISC acts on the zero-field triplet states, given by the eigenstates $|X\rangle$, $|Y\rangle$ and $|Z\rangle$ of the Hamiltonian (where X,Y and Z are the principal axes of the ZFS tensor $\boldsymbol{D}$), which are split due the ZFS parameters $D$ and $E$.[32, 33] The ISC rate depends on the difference in the nature of the singlet and triplet excited states involved in this process, whereby the population is spin selective, depending on the symmetries of the excited singlet and triplet wave functions.[33, 34, 59, 60] In the studied materials,

SOC-driven ISC leads to a relative population $p_i$ of zero-field triplet sublevels, in, e.g. neat Y6 of [$p_z$, $p_y$, $p_x$] = [0, 0.66, 0.34] (see Section 6 and Table S7 in SI). In a magnetic field, the population of the corresponding high-field sublevels $|T_+\rangle$, $|T_0\rangle$, $|T_-\rangle$ can be derived from the zero-field population, which depends on the orientation of the ZFS tensor with respect to the external magnetic field (here shown for $\vec{B}||Z$, further details in Figure S13).[33] In this projection, the $|T_+\rangle$ and $|T_-\rangle$ states are more populated than $|T_0\rangle$, with resonant microwave irradiation driving this imbalance towards equal distributions. Together with the asymmetric splitting due to ZFS, microwave emission (*e*) at lower fields (negative intensity) and enhanced absorption (*a*) at higher fields (positive intensity) can be detected with trEPR.[34] In disordered organic samples, transitions from all orientations of ***D*** with respect to $\vec{B}$ are superimposed as a characteristic spectrum with *eeeaaa* pattern, which is indicative for triplet excitons populated by SOC-driven ISC (Figure S13).[33] Returning to Figure 2a, the SOC-driven ISC pattern of Y6 triplets is also present in the blends of PM6:Y6 and PM7:Y6 with similar zero-field population of triplet sublevels as neat Y6 (Table S7), indicating ISC on the NFA.

## 3. Influence of Halogenation on Excited State Kinetics

### 3.1 Hole Transfer in (Non-) Halogenated PBDB-T:Y-Series Studied by Transient Absorption

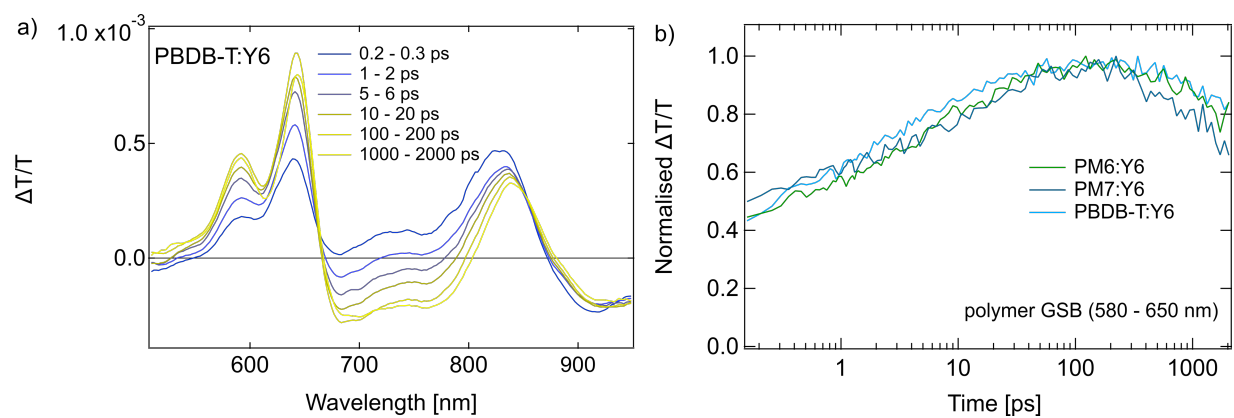

**Figure 4.** Transient absorption (TA) measurements. a) TA spectra at different time delays for PBDB-T:Y6. b) Hole transfer kinetics extracted from polymer ground state bleach GSB (580 – 650 nm) for PM6:Y6, PM7:Y6 and PBDB-T:Y6. In all samples, hole transfer takes place on comparable timescales. All TA measurements used a pump pulse wavelength of 800 nm for selective excitation of Y6 and were performed at *T* = 293 K.

**Figure 4** shows TA spectra and hole transfer kinetics for Y6 blends with excitation wavelength of 800 nm for selective excitation of the NFA Y6 (blends with Y7 discussed in Section 3 of SI). The low fluence of ~0.6 µJ cm$^{-2}$ used in the TA measurements on the blends prevents excess bimolecular recombination in the first few nanoseconds from affecting the determination of the hole transfer timescales.

**Figure 4a** shows the time dependent TA spectra of PBDB-T:Y6; TA spectra for PM6:Y6 and PM7:Y6 are shown in Figure S1. The Y6 ground state bleach (GSB) centered at 830 nm and photo-induced absorption (PIA) of Y6 singlet excitons at 910 nm are observed immediately after optical excitation.[44, 61] In addition, a clear GSB at 580 - 650 nm emerges in the TA spectra, consistent with the absorption spectrum of the polymer.[62, 63] Since the excitation pump does not generate excitons on PBDB-T (PM6, PM7 respectively), the appearance of the polymer GSB indicates dissociation of Y6 excitons via hole transfer at the D:A interface. The presence of the polymer GSB already at 0.2 ps in all blends suggests that some ultrafast hole transfer occurs.[61, 64] However, as the Y6 GSB falls, likely due to some spectral overlap with the negative sign PIAs of Y6 electron and PM6 hole polarons forming at 780 nm and 920 nm[65], respectively, the polymer GSB continues to increase, peaking around 100 ps. This indicates that a significant population of the NFA excitons require additional time to dissociate. On similar timescales, another broad negative band grows in between 680 and 730 nm. This has previously been assigned to the electro-absorption of the donor polymer PM6, indicating the separation of bound interfacial CT states into free charges.[3, 61, 65] However, whilst all blends with the NFA Y6 show a considerable hole transfer yield, it decreases upon halogenation of the donor in these blends (Figure S1) and with NFA Y7 (Figure S4), further discussed below and in Sections 2 and 3 of SI.

**Figure 4b** shows the normalized hole transfer kinetics from the polymer GSB between 580 and 650 nm. The hole transfer rate appears to be comparable for all donor combinations (PBDB-T, PM6 and PM7) with Y6. In addition to the ultrafast component already visible at 0.2 ps, the hole transfer yield increases up to a peak at 100 ps, indicating that the hole transfer is completed. Donor combinations with NFA Y7 (Figure S4) also show very similar hole transfer kinetics to Y6 blends. The slower hole transfer rate component is often attributed to diffusion limitations for excitons created far from the D:A interface.[66, 67] Zhong *et al.* performed measurements on 5:1 polymer:NFA blends to disentangle the ultrafast component, determined by intrinsic CT rates, from the morphology dependent exciton diffusion rates.[64] However, the studied blends show similar kinetics in both components, reflecting that neither the small energetic shift in HOMO level (Figure S17) nor probable morphology differences upon halogenation have a significant impact on the hole transfer rates, which has been found to be largely governed by the interfacial morphology in PM6:Y6.[61] Interestingly, despite initially generating a similar number of excitations on Y6 (as seen by the equal intensity of the Y6 GSB at 0.2 ps in all blends), the intensity of the polymer GSB reaches the highest value in PBDB-T, followed by PM6 and then PM7 (Figure S1). This indicates a difference in the absolute hole transfer yield, implying that a smaller number of Y6 excitons reaches the D:A interface for hole transfer.

## 3.2 SOC-driven ISC in (Non-) Halogenated Y-Series by Quantum-Chemical Calculations

Singlet excitons formed upon photoexcitation which do not reach the D:A interface for efficient hole transfer may eventually undergo ISC to the respective triplet states. We performed QC calculations to assess possible differences in the ISC rates due to halogenation. In the assessment of the energetic landscape of the investigated donor polymers and the NFA Y-series, singlet and triplet vertical excitation energies were computed by means of *screened* range-separated hybrid (SRSH) time-dependent DFT calculations, performed within the Tamm-Dancoff approximation (TDA) (see SI for further details and Table S2).[68, 69] In order to obtain triplet energies, the theoretical singlet-triplet energy gap $\Delta E_{ST}$ was subtracted from experimental singlet energies (Table S2). Then, we focused on the calculation of ISC rates $k_{ISC}$ for the NFA Y-series and we resorted to the semi-classical Marcus-Levich-Jortner expression, which treats high-frequency intramolecular vibrational modes in a quantum-mechanical fashion:

$$k_{\text{ISC}} = \frac{2\pi}{\hbar} |\langle T_x|\hat{H}_{SOC}|S_1\rangle|^2 \sqrt{\frac{1}{4\pi\lambda_s k_B T}} \times \sum_n \left\{ \exp(-S)\frac{S^n}{n!} \times exp\left[-\frac{\left(-\Delta E^0_{S1-Tx} + \lambda_s + n\hbar\omega\right)^2}{4\lambda_s k_B T}\right]\right\}$$

Here, $|\langle T_x|\hat{H}_{SOC}|S_1\rangle|$ displays the SOC matrix element involved in the ISC process from $|S_1\rangle$ to the manifold of the triplet states $|T_x\rangle$, whose energy differences are given by $\Delta E^0_{S1-Tx}$; $\lambda_s$ is the external reorganization energy (set in a range between 0.05 and 0.2 eV); $\hbar$ is the reduced Planck's constant; $k_B$ is the Boltzmann constant; and $T$ is the temperature (set at 298 K). $\hbar\omega$ is the energy of an effective high-frequency intramolecular vibrational mode (0.15 eV of a carbon-carbon stretching) that assists the ISC and $S = \lambda_i/(\hbar\omega)$ is the Huang-Rhys factor, which is a measure of the electron-phonon coupling assisting the ISC process and $\lambda_i$ the internal reorganization energy.

The nature of $|S_1\rangle$ excited state of Y6 and Y7 is very similar to that of $|T_1\rangle$ excited state which makes SOC between these excited states negligible according to the El-Sayed's rule.[70, 71] For instance, in Y6 both $|S_1\rangle$ and $|T_1\rangle$ show a dominant HOMO-to-LUMO transition (98% for $|S_1\rangle$ and 90% for $|T_1\rangle$, in agreement with literature[72], see Figure S7) leading to a very small SOC value (< 0.05 cm$^{-1}$). In contrast, the $|T_2\rangle$ excited state (which is energetically also below $|S_1\rangle$) shows a contribution of 75% of HOMO-to-LUMO+1 transition, involving a different localization of the electron wave function (see Figure S7) in comparison to $|S_1\rangle$. Consequently, SOC between $|S_1\rangle$ and $|T_2\rangle$ is much larger (on the order 0.1 cm$^{-1}$) compared to the SOC between $|S_1\rangle$ and $|T_1\rangle$, enabling singlet-triplet conversion by ISC on the NFAs.

Moreover, to study the influence of halogenation, we compared Y6 and Y7 together with the non-halogenated version Y5. The same natural transition orbitals (NTOs) were found in all three cases, meaning that the relevant low-lying excited states (namely $|S_1\rangle$ and $|T_2\rangle$) have the very same nature and halogenation does not play a key role here. In fact, the electron density on the halogen atoms in the NTO is negligible (Figure S7), resulting in the SOC matrix element being almost insensitive to halogenation. As the respective $\Delta E_{ST}$ is also not changing considerably (Table S2) and, additionally, has been shown to have a minor influence on the ISC rate, the SOC-driven ISC kinetics from $|S_1\rangle$ via $|T_2\rangle$ can be assumed to be comparable for all (non-) halogenated acceptors.

## 4. Results and Discussion

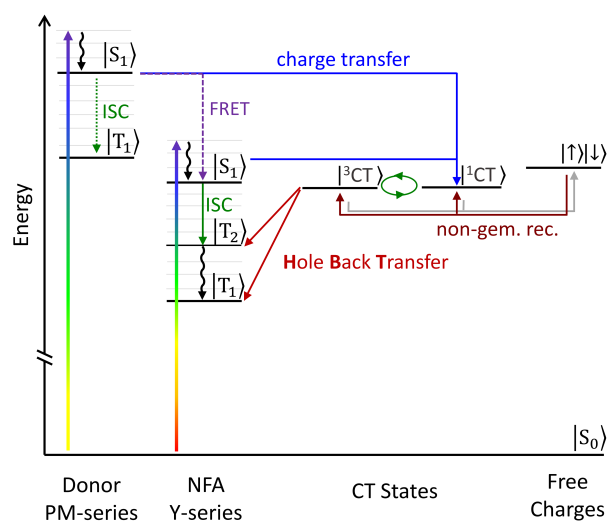

**Figure 5.** Energy diagram of the studied OPV blends. Optical excitation of donor or acceptor singlet is followed by interfacial charge transfer, FRET or direct ISC. NFA triplet states ($|T_1\rangle$, $|T_2\rangle$) are energetically below the intermolecular CT states ($|^1CT\rangle$, $|^3CT\rangle$) and therefore represent energetic trap states. NFA triplet states are populated by ISC from the NFA $|S_1\rangle$ state or by HBT from $|^3CT\rangle$, which itself is populated by non-geminate recombination of free charges.

**Figure 5** shows the Jablonski diagram of the studied PBDB-T:Y-series blends. All energies and observed triplet formation mechanisms are summarized in Table S1. Given the low HOMO offset in all investigated blends, the low-lying NFA singlet states possess a small energetic offset to CT states, allowing small energy losses in comparison to fullerene-based solar cells.[5, 44] However, given a large $\Delta E_{ST}$ of molecular states (≈0.5 – 0.6 eV), the small singlet offset results in the lowest NFA triplet being energetically below the CT states. Such a large $\Delta E_{ST}$ prevents efficient triplet harvesting mechanisms, such as thermal upconversion, to directly return to the interfacial CT states and to yield photocurrent. Without further interactions, triplets are instead energetically trapped and, given the small first order decay rates to singlet ground states, can be detected by both PLDMR and trEPR.[3, 29] The fact that triplet excitons were detected in all PBDB-T:Y-series blends should draw attention to the role of the triplets in the solar cell power loss that results as a direct consequence of the energetic reduction of the HOMO offset.[3] However, to be able to control the population of these triplet states, we studied in detail the photophysical processes and pathway differences between the state-of-the-art (non-) halogenated donor and acceptor blends, which surprisingly highlights the non-halogenated donor PBDB-T. In the following, we discuss the different pathways affecting the population of the triplet states (direct ISC from $|S_1\rangle$ states upon photoexcitation as well as HBT from $|^3CT\rangle$ states formed via either geminate or non-geminate recombination) and the impact of halogenation on the major kinetic rates.

**4.1 Triplet Formation by Geminate and Non-Geminate Pathways**

Triplet excitons were detected in all PBDB-T:Y-series blends on the NFA triplet states via PLDMR (Figure 2 and Section 5 in SI). In trEPR (Figure 3), probing predominantly geminate excitons, the polarization patterns for the detected triplets with donors PM6 and PM7 are consistent with polarization for SOC-driven ISC triplets (*eeeaaa*). These direct ISC-driven triplet states can be clearly distinguished from other triplet formation processes, such as HBT from $|^3CT\rangle$ states formed via geminate recombination as observed to be a dominant process in polymer:fullerene blends.[16, 29] However, in the investigated NFA-based blends, no indication of geminate HBT could be observed, ruling out this triplet formation mechanism, consistent with reports in the literature.[3] The comparable zero-field populations of the blends and neat Y6 (see Section 6 and Table S7 in SI) indicate that ISC takes place on the NFA, e.g. in the center of NFA domains larger than the exciton diffusion length away from the D:A interface. QC calculations revealed the SOC-driven ISC rate is increased by the presence of higher-lying triplet states with larger variation in their nature with respect to $|S_1\rangle$, enhancing the SOC matrix element. As for the donor, ultrafast electron transfer quenches the singlet excitation in the sub-picosecond timescale, with the possibility of Förster resonance energy transfer (FRET) to the NFA as another deactivation channel, suggesting the yield of excitons undergoing ISC on the donor to be negligible (dotted arrow in Figure 5).[11, 61, 67, 73] Furthermore, donor triplet levels are energetically above the CT and NFA states and would be able to undergo charge or Dexter energy transfer, leading to a minor impact on efficiency loss.

The NFA triplet yield by SOC-driven ISC is assumed to be low in these OSCs, as demonstrated by a weak trEPR signal in the spin-coated substrates (Figure S2). However, we additionally confirmed HBT from $|^3CT\rangle$ states formed via non-geminate recombination of free charge carriers, visible in PLDMR spectra of all blends. The origin lies in the recombination of free charge carriers with uncorrelated spin orientation that lead to the formation of 25% CT singlets and 75% CT triplets.[74] The latter can relax spin-allowed to the lowest molecular triplet states via non-geminate HBT (dark red arrows in Figure 5).[17] Thereby, excitons initially occupy the three triplet sublevels equally according to spin statistics, but will gain a small spin polarization due to a combination of spin-lattice relaxation and unequal triplet sublevel decays. Continuous optical excitation in PLDMR further enhances the spin polarization due to accumulation-driven spin-dependent annihilation processes. In combination with the optical detection, this improves the sensitivity to non-geminate triplet excitons in contrast to trEPR. Gillett *et al.* showed with TA spectroscopy that in the blend PM6:Y6 around 90% of the non-geminately recombined CT states undergo HBT to the NFA triplet.[3] In contrast, the ISC yield in neat Y6 is around 3%, further reduced in the blends due to a significant charge transfer of 90% upon photoexcitation.[3] Regarding the similarity of ISC and hole transfer rates discussed below, we assume similar amounts of triplet excitons from non-geminate recombination in all PBDB-T:Y-series blends, as detected via PLDMR (Figure 2). Furthermore, trPLDMR of PM6:Y6 blend films shows a similar temporal response of the Y6 triplets and the CT states, supporting the population by non-geminate HBT, while Y6 triplets in neat film possess a different time response in agreement with triplets predominantly populated by ISC (Figure S10). Since the rate of non-geminate HBT is shown to increase with temperature, the presence of non-geminate HBT triplet excitons even at 10 K highlights the importance of this loss channel for solar cells operated at ambient temperatures.[3] For further evidence of dominant non-geminate recombination at room temperature, we measured TA at 293 K in the infrared region, where the formation of the NFA triplet PIA is observable (Figure S3). The strong fluence-dependence of the Y6 triplet generation in this spectral region attributes the population to a dominating bimolecular/non-geminate recombination in all (non-) halogenated blends, as already identified by PLDMR at low temperatures. These triplet recombinations have been shown to result in a non-radiative voltage loss of up to 60 meV in operational solar cells, with our study implying a similar degree with respect to new donor or acceptor halogenations.[3]

**4.2 Impact of Halogenation on Triplet Formation**

The number of excitons undergoing the above-mentioned processes additionally depends on the kinetics of the system, whereby the two most important rates affecting the population of triplet states upon initial excitation of the singlet states are hole transfer and ISC rates. As differences in hole transfer kinetics would give the singlet excitons more/less time to undergo ISC, we studied the impact of halogenation on both rate constants. Regarding the halogenation of the acceptor, one could invoke an increase in SOC due to heavier halogen atoms, boosting ISC as has already been shown in literature for bromine or iodine in TADF molecules.[75] However, QC calculations on NFA Y-series reveal that, starting from the non-halogenated acceptor Y5, the ISC rate constant does not change significantly upon fluorination and chlorination in these molecules due to the absence of significant weight on the halogen atoms (see Table S4, Figure S7 for Y6 and Figure S8 for the PM6). As for the hole transfer kinetics, halogenation of PBDB-T lowers the donor HOMO energy level (Figure S17), leading to increased $V_{OC}$ in solar cells with PM6 and PM7.[62, 76] The smaller energy offset is desirable to minimize energy losses but one could consider an effect on the driving force due to a smaller HOMO offset.[20,

[67] However, TA measurements (Figure 4b) showed a comparable hole transfer rate constant with the three different (non-) halogenated donors and with both NFAs (Figure S4). This outcome is not surprising, as the overall hole transfer rate has been shown not being sensitive to the HOMO offset, especially in low-offset OSC: while the 'intrinsic' hole transfer rate of NFA excitons generated in close proximity to the D:A interface occurs on sub-ps timescales, the overall hole transfer process takes place over timescales of tens of picoseconds, depending on exciton diffusion in the bulk morphology.[64] Furthermore, Ma *et al.* showed that hole transfer rates for PM6:Y6 are only weakly affected by temperature in the range from 15 – 300 K, allowing the comparison with the low-temperature EPR measurements.[73]

However, while the triplet population kinetics upon initial singlet excitation are shown to be comparable for (non-) halogenated donor and acceptor combinations, the number of excitons undergoing these processes is determined to be different. Comparing TA and trEPR for all Y6 blends with substrates prepared in the same way according to the optimized recipe (Figure S1 and Figure S2), the hole transfer yield decreases upon halogenation, while trEPR data show a detectable ISC yield in PM6:Y6 and PM7:Y6. This result surprisingly suggests a positive effect of using the non-halogenated donor PBDB-T by revealing a reduced triplet exciton yield. As already mentioned, halogenation also impacts the morphology by, e.g. increasing domain sizes due to fluorination or chlorination as a result of a reduced miscibility.[77-79] These increased domain sizes are suggested to lead to a decrease in the number of excitons reaching the D:A interface and instead undergoing ISC in blends with halogenated donor PM6 and PM7. Indeed, Eastham *et al.* investigated different donor materials with NFA ITIC, demonstrating that efficient hole transfer depends strongly on the blend morphology rather than the energy level alignment.[80] Domain sizes in the range of the exciton diffusion length are normally beneficial for dissociation of charge carriers, improving short-circuit current density and fill factor of the solar cell.[77, 78] However, we showed by the analysis of the spin physics in these systems, that triplet recombination pathways can be unfavourably affected. Thus, the key point for a good performing OSC is matching a good trade-off in efficient photophysical processes and beneficial bulk morphology to reduce charge generation losses, as in PBDB-T:Y6, but also beneficial subsequent exciton dissociation, as in PM6:Y6 and PM7:Y6.[81]

There are different design strategies for next-generation OSCs to engineering out the triplet loss pathways: One approach is to reduce the HBT rate to molecular triplet states by hybridization between CT and local excitons or to outcompete triplet recombination by fast charge separation.[3, 82] Another approach is the combination of state-of-the-art OPV and OLED research by designing the second OSC component with small singlet-triplet gap, e.g., intramolecular TADF emitters. This combination would generally prevent energetically trapped triplet excitons by thermal reactivation, while simultaneously preserve the small voltage losses.

## 5. Conclusion

In this work, we applied triplet spin-sensitive techniques (PLDMR, trEPR) together with TA and QC calculations to reveal non-radiative loss processes by triplet excitons. Using the strengths of these complementary methods, we were able to detect and distinguish various triplet excitons present in the investigated blends and draw a comprehensive picture of their generation pathways. We thereby studied different state-of-the-art donor:acceptor combinations with the (non-) halogenated polymers PBDB-T, PM6, PM7 and the NFAs Y6 and Y7 and detected triplet excitons in all blends. These excitons are energetically trapped on the low-lying NFA triplet states, where they are unable to contribute to OSC performance due to a high energetic gap to CT states. While the major contribution observed in all blends is non-geminate HBT, blends with halogenated donors possess an additional triplet population pathway through SOC-driven ISC on the NFA. The impact of halogenation on the rates affecting the population of triplet states after excitation of the NFA singlet states, in particular ISC and hole transfer, is comparable for all combinations. Thus, shifting of HOMO levels or the presence of heavier halogen atoms have a minor influence on excited states kinetics. However, the increased triplet exciton yield by ISC with best performing PM6 and PM7 indicates an incomplete charge carrier generation, suggesting an adverse impact of domain aggregation through halogenation. While the benefits of halogenation still prevail in terms of overall device efficiency, we uncovered triplet excitons and associated pathways in all these (non-) halogenated blends which clearly shows that there is untapped potential for reaching the 20% milestone.

## 6. Experimental Section

As donor materials, poly[[4,8-bis[5-(2-ethylhexyl)-2-thienyl]benzo[1,2-*b*:4,5-*b'*]dithiophene-2,6-diyl]-2,5-thiophenediyl[5,7-bis(2-ethylhexyl)-4,8-dioxo-4*H*,8*H*-benzo[1,2-*c*:4,5-*c'*]dithiophene-1,3-diyl]]polymer (PBDB-T), poly[[4,8-bis[5-(2-ethylhexyl)-4-fluoro-2-thienyl]benzo[1,2-*b*:4,5-*b'*]dithiophene-2,6-diyl]-2,5-thiophenediyl[5,7-bis(2-ethylhexyl)-4,8-dioxo-4*H*,8*H*-benzo[1,2-*c*:4,5-*c'*]dithiophene-1,3-diyl]-2,5-thiophenediyl] (PBDB-T-2F or PM6) and poly[[4,8-bis[5-(2-ethylhexyl)-4-chloro-2-thienyl]benzo[1,2-*b*:4,5-*b'*]dithiophene-2,6-diyl]-2,5-thiophenediyl[5,7-bis(2-ethylhexyl)-4,8-dioxo-4*H*,8*H*-benzo[1,2-*c*:4,5-*c'*]dithiophene-1,3-diyl]-2,5-thiophenediyl] (PBDB-T-2Cl or PM7) were used. NFAs represent 2,2'-((2Z,2'Z)-((12,13-bis(2-ethylhexyl)-3,9-diundecyl-12,13-dihydro-[1,2,5]thiadiazolo[3,4-e]thieno[2'',3'':4',5']thieno[2',3':4,5]pyrrolo[3,2-g]thieno[2',3':4,5]thieno[3,2-b]indole-2,10-diyl)bis(methanylylidene))bis(5,6-difluoro-3-oxo-2,3-dihydro-1H-indene-2,1-diylidene))dimalononitrile (BTP-4F or Y6) and 2,2'-((2Z,2'Z)-((12,13-bis(2-ethylhexyl)-3,9-diundecyl-12,13-dihydro-[1,2,5]thiadiazolo[3,4-e]thieno[2'',3'':4',5']thieno[2',3':4,5]pyrrolo[3,2-g]thieno[2',3':4,5]thieno[3,2-b]indole-2,10-diyl)-

bis(methanylylidene))bis(5,6-dichloro-3-oxo-2,3-dihydro-1H-indene-2,1-diylidene))dimalononitrile (BTP-4Cl or Y7). All materials were purchased from 1-materials or Sigma Aldrich.

Substrates were prepared from solution according the optimized recipe (18 mg/ml, 1:1.2 polymer:NFA, dissolved in CF with 1-chloronaphthalene 0.5 v/v%)[37, 38], which was spin-coated with 3000 rpm for 30 s on substrates (10 mm x 10 mm for TA, 20 mm x 20 mm for PLDMR/trEPR) and annealed at 110°C for 10 min. TA samples were encapsulated under nitrogen atmosphere with 20 mm x 20 mm x 0.2 mm cover glass. Samples for PLDMR and trEPR were cut into strips of 2 mm width, whereby 10 strips were placed in an EPR quartz tube (Wilmad O.D. 4mm). The sample tubes were sealed under inert helium atmosphere. For EPR measurements on drop-cast samples, materials and blends were dissolved in chlorobenzene (5 mg/ml for PLDMR, 20 mg/ml for trEPR) and ~100 µl were poured into an EPR tube. The solvent was then evaporated by vacuum pumping, creating a thin film on the EPR tube wall. The sample tubes were also sealed under inert helium atmosphere.

PLDMR and trEPR experiments were carried out with a modified X-band spectrometer (Bruker E300) equipped with a continuous-flow helium cryostat (Oxford ESR 900) and a microwave cavity (Bruker ER4104OR, 9.43 GHz) with optical access. All measurements were performed at $T$ = 10 K.

For PLDMR, microwaves were generated with a microwave signal generator (Anritsu MG3694C), amplified to 3 W (microsemi) and guided into the cavity. Optical irradiation was performed with a 473 nm continuous wave laser (Cobolt). PL was detected with a silicon photodiode (Hamamatsu S2281) on the opposite opening of the cavity, using a 561 nm longpass filter to reject the excitation light. The PL signal was amplified by a current/voltage amplifier (Femto DHPCA-100) and recorded by lock-in detector (Ametek SR 7230), referenced by on-off-modulating the microwaves with 517 Hz.

For trEPR, pulsed optical excitation was performed with a Nd:YAG laser (Continuum Minilite II) with 532 nm; pulse length of 5 ns; 15 Hz repetition rate; 2 mJ per pulse. Microwaves were generated and detected with a microwave bridge (Bruker ER047MRP). Measurements were performed with 20 dB attenuation (2 mW). A voltage amplifier (FEMTO DHPVA-200) and a digitizer card (GaGe Razor Express 1642 CompuScope) were used for transient recording. The time resolution is limited to ~100 ns by the cavity Q factor of around 2800. By sweeping the magnetic field, two-dimensional data sets are recorded, where trEPR spectra are averaged from 0.5 – 1.5 µs after laser excitation.

For TA, a setup powered by a Yb amplifier (PHAROS, Light Conversion), operating at 38 kHz and generating 200 fs pulses centred at 1030 nm with an output of 14.5 W, was used. The ~200 fs pump pulse was provided by an optical parametric amplifier (Light Conversion ORPHEUS). The probe was provided by a white light supercontinuum generated in a YAG crystal from a small amount of the 1030 nm fundamental. After passing through the sample, the probe is imaged using a Si photodiode array (Stresing S11490).

QC calculations were carried out by using time-dependent DFT calculations within the Tamm-Dancoff approximation (TDA). For these calculations, optimally-tuned *screened* range-separated hybrid (SRSH) LC-ωhPBE/6-311G(d,p) level of theory was employed, where the scalar dielectric constant was set at 4.5. In addition, SOC matrix elements were computed in the Brett-Pauli spin-orbit Hamiltonian framework as implemented in the PySOC code[83], performing the calculations at the ωB97X-D/Def2-TZVP level on the NFA Y-series $S_1$ optimized structure. All the calculations were carried out with the Gaussian16 suite[84]. Further details can be found in SI.

**Acknowledgements**

J.G., V.D. and A.S. acknowledge support by the Deutsche Forschungsgemeinschaft (DFG, German Research Foundation) within the Research Training School "Molecular biradicals: Structure, properties and reactivity" (GRK2112). M.K., G.L., V.D. and A.S. acknowledge EU H2020 for funding through the Grant SEPOMO (Marie Skłodowska-Curie Grant Agreement 722651). Computational resources were provided by the Consortium des Équipements de Calcul Intensif (CÉCI), funded by the Fonds de la Recherche Scientifiques de Belgique (F.R.S.-FNRS) under Grant No. 2.5020.11, as well as the Tier-1 supercomputer of the Fedération Wallonie-Bruxelles, infrastructure funded by the Walloon Region under Grant Agreement No. 1117545. G.L. and Y.O. acknowledges funding by the Fonds de la Recherche Scientifique-FNRS under Grant No. F.4534.21 (MIS-IMAGINE). A.J.G. thanks the Leverhulme Trust for an Early Career Fellowship (ECF-2022-445). J.G. acknowledges support from the EPSRC (EP/W017091/1). The authors thank David Beljonne for fruitful discussion.

**Data Availability**

All data needed to evaluate the conclusions in the paper are present in the paper and/or the Supplementary Information.

**Conflicts of interest**

There are no conflicts to declare.

**Authors contributions**

J.G., B.S., and S.L. performed the magnetic resonance measurements and evaluated the data. G.L. and Y.O. performed the calculations. A.J.G. performed the transient absorption measurements. J.G. and A.S. wrote the manuscript, which all authors discussed and commented on.

**Supplementary Materials**

Supporting Information contains supplementary explanation and data, including additional TA spectra, computational details, additional PLDMR/trEPR spectra and triplet simulation parameters. SI includes Figures S1 – S20, Table S1 – S7.

**ORCID**

Jeannine Grüne: 0000-0003-3579-0455

Giacomo Londi: 0000-0001-7777-9161

Alexander J. Gillett: 0000-0001-7572-7333

Yoann Olivier: 0000-0003-2193-1536

Vladimir Dyakonov: 0000-0001-8725-9573

Andreas Sperlich: 0000-0002-0850-6757

# SUPPORTING INFORMATION

## for

**Triplet Excitons and Associated Efficiency-Limiting Pathways in Organic Solar Cell Blends Based on (Non-) Halogenated PBDB-T and Y-Series**

Jeannine Grüne[1,2]*, Giacomo Londi[3], Alexander J. Gillett[2], Basil Stähly[1], Sebastian Lulei[1], Maria Kotova[1], Yoann Olivier[3], Vladimir Dyakonov[1], and Andreas Sperlich[1]*

[1]Experimental Physics 6, Julius Maximilian University of Würzburg, Am Hubland, 97074 Würzburg, Germany

[2]Cavendish Laboratory, University of Cambridge, JJ Thomson Avenue, Cambridge, UK

[3]Laboratory for Computational Modeling of Functional Materials, Namur Institute of Structured Matter, University of Namur, Rue de Bruxelles, 61, 5000 Namur, Belgium

* jg2082@cam.ac.uk, sperlich@physik.uni-wuerzburg.de

**Contents**

## 1. Energy Levels

| Material | $S_1$ [eV] | $\Delta E_{ST}$ [eV] | $T_1, T_2$ [eV] | CT [eV] | Triplets | Mechanism | Refs. for $S_1$, CT |
|---|---|---|---|---|---|---|---|
| PBDB-T | 1.85 | 0.40 | 1.45 | | | ISC | [1, 2] |
| PM6 | 1.92 | 0.41 | 1.51 | | | ISC | [3, 4] |
| PM7 | 1.92 | 0.41 | 1.51 | | | ISC | [5] |
| Y6 | 1.39 | 0.56, 0.35 | 0.83, 1.04 | | | ISC | [2, 3, 6-10] |
| Y7 | 1.40 | 0.55, 0.34 | 0.85, 1.06 | | | ISC | [11] |
| PBDB-T:Y6 | | | | 1.35 | Y6 | HBT (ng) | [12] |
| PM6:Y6 | | | | 1.37 | Y6 | ISC, HBT (ng) | [2, 7, 10, 13, 14] |
| PM7:Y6 | | | | 1.37 | Y6 | ISC, HBT (ng) | [15] |
| PBDB-T:Y7 | | | | 1.36 | Y7 | HBT (ng) | [15] |
| PM6:Y7 | | | | 1.38 | Y7 | ISC, HBT (ng) | [12] |
| PM7:Y7 | | | | 1.38 | Y7 | ISC, HBT (ng) | n.a. |

**Table S1.** Energy values for donor polymers PBDB-T, PM6, PM7 and acceptors Y6, Y7. $S_1$ and CT energies of experimentally determined values are taken from given references (averaged values). Triplet energies are determined by subtraction of calculated $\Delta E_{ST}$ from $S_1$. CT singlet and triplet energies are treated as isoenergetic. Triplets in blends are localized on acceptor molecules populated by different formation mechanisms (ng = non-geminate). The CT energy for PM7:Y7 is assumed to be identical with PM6:Y6 because PM6 and PM7 have identical energy levels.

## 2. Additional trEPR and TA Data for Y6 Blended Spin-Coated Substrates

PM6:Y6 solar cells with the optimized recipe have a spin-coated active layer of 14 – 18 mg/ml PM6:Y6 blend in a ratio of 1:1.2 dissolved in chloroform (CF) with 0.5% v/v 1-chloronaphtalene (1-CN).[16, 17] The solution is spin-coated with 3000 rpm for 30 sec and subsequently annealed at 110° for 10 min. The blends for the other donor:acceptor combinations are prepared with the same recipe. While TA measurements are performed on glass substrates with size of 10x10 mm (encapsulated with 20x20 mm cover slip of 0.2 mm thickness), for EPR measurements, 20x20 mm cover slips were cut into strips of 2 mm width whereby 10 strips were placed into an EPR tube.

### TA Measurements in the visible region (hole transfer)

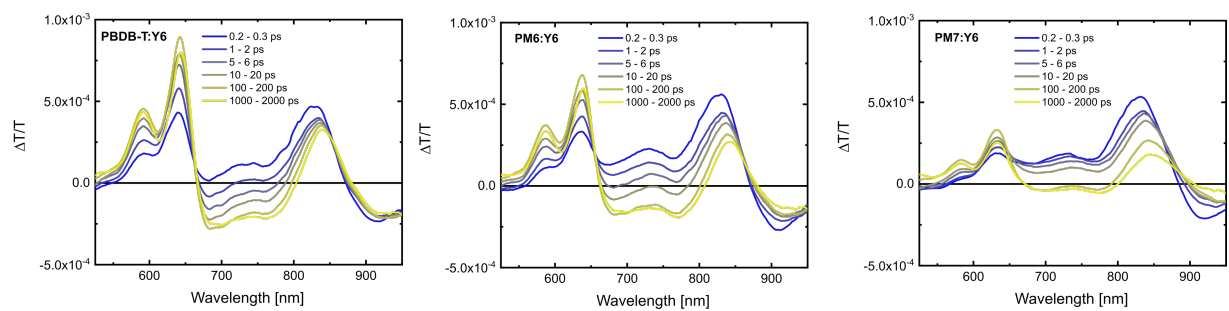

**Figure S1** Transient absorption (TA) spectra of PBDB-T:Y6, PM6:Y6 and PM7:Y6. Excitation at 800 nm; fluence of ~0.6 µJ cm$^{-2}$.

Figure S1 shows the TA measurements for PBDB-T:Y6, PM6:Y6 and PM7:Y6. All samples show a comparable Y6 GSB at 840 nm, indicating a comparable number of excitations created on Y6. While a clear polymer GSB is visible in PBDB-T:Y6 and PM6:Y6 already after 0.2 ps, the polymer GSB at 0.2 ps PM7:Y6 is very weakly pronounced. Comparing the intensity of polymer GSB when hole transfer is completed at around 100 ps, the hole transfer yield in PBDB-T:Y6 is the highest, followed by PM6:Y6 and PM7:Y6. In the same trend, the Y6 GSB is decaying slowest in PBDB-T:Y6, followed by PM6:Y6 and PM7:Y6. While the hole transfer kinetics are comparable in all three blends, the faster Y6 GSB decay is most likely due to undissociated Y6 $S_1$ excitons recombining to the ground state or undergoing ISC.

**trEPR Measurements**

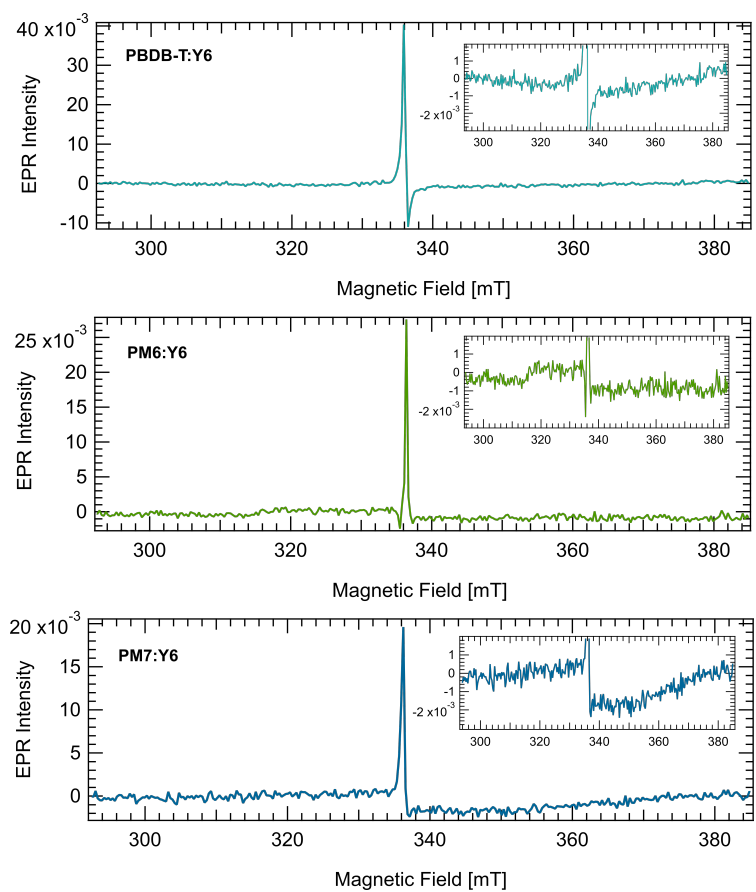

**Figure S2.** trEPR measurements on spin-coated substrates for Y6 blends. All blends possess an intense CT peak whereby PM6:Y6 and PM7:Y6 show a small ISC yield.

Figure S2 shows the trEPR measurements on spin-coated substrates for Y6 blends. As for the dropcast samples in Figure 2, an ISC yield is visible for PM6:Y6 and PM7:Y6, as already shown in literature for spin-coated PM6:Y6 substrates.[14] In contrast to the dropcast samples of Figure 2, the molecular triplet signal is smaller. On the one hand, the smaller amount of materials on the substrates and, on the other hand, the increased hole transfer yield in spin-coated films reduces the molecular signal. The exact pattern cannot be determined from this signal-to-noise ratio, but it fits the emission/absorption in SOC-driven ISC as already observed in the dropcast samples. Moreover, the width corresponds to triplet excitons on the NFA. PBDB-T:Y6 shows the same trend in ISC yield as the dropcast samples with negligible trEPR signal, which is consistent with the TA measurements. Nevertheless, the ISC yield in the other two blends is also very low, so the competing process to hole transfer from Figure S1 is most probably the $S_1$ decay to the ground state. Moreover, PBDB-T:Y6 shows the highest CT peak, indicating the highest fraction of CT states, while the CT peak decreases in PM7:Y6.

**TA measurements in the infrared region (Y6 triplet excitons)**

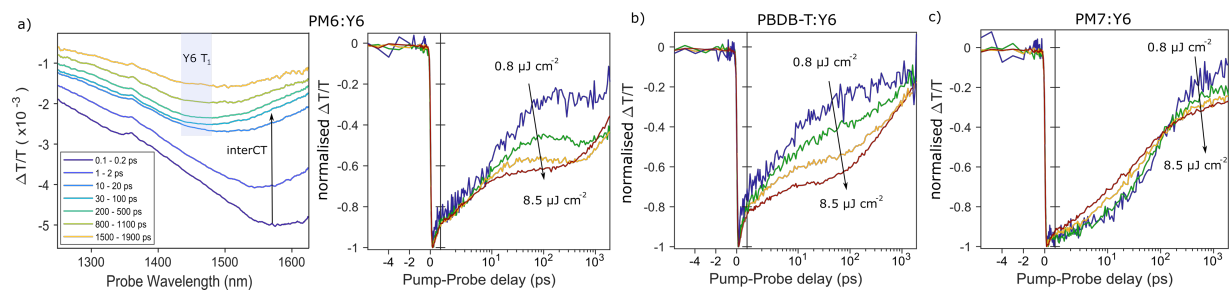

**Figure S3.** Transient absorption (TA) measurements. a) TA spectra at different time delays for PM6:Y6 in the infrared region and kinetics extracted at the Y6 triplet PIA (~1450 nm). The fluence dependence (0.8 µJ cm$^{-2}$ to 8.5 µJ cm$^{-2}$) confirms triplet excitons formed via bimolecular/non-geminate recombination, dominating at room temperature. b) Kinetics extracted at the Y6 triplet PIA for PBDB-T:Y6 and c) for PM7:Y6 (the later appearance of the triplet feature in PM7:Y6 is caused by a slower interCT PIA decay, spectrally overlapping with the Y6 $T_1$).

To compare the triplet generation at low temperatures in trEPR and PLDMR to room temperature measurements, we performed TA spectroscopy with probing the infrared region (1250 – 1625 nm) to investigate the NFA triplet population. Explained exemplary for PM6:Y6 (Figure S3a), we observe a PIA feature on early time scales (0.1 – 0.2 ps after excitation) at 1550 nm, corresponding to intermolecular charge transfer excitation (interCT) between neighboring Y6 molecules.[9] After the hole transfer develops over the first hundred picoseconds (see Figure 4b in the main manuscript), this PIA decreases and a new PIA at 1450 nm rises after 30 ps. This new PIA (Figure S3a, blue box) can be attributed to the Y6 triplet and decays over long timescales (measured up to 1900 ps). The strong fluence-dependence of the Y6 triplet formation in this spectral region (Figure S3a, right) attributes the population to a dominating bimolecular/non-geminate recombination, as already identified by PLDMR at low temperatures. Figures S3b,c show the kinetics for the Y6 triplet region also for PBDB-T:Y6 and PM7:Y6, with the fluence-dependent kinetics (at later times in PM7:Y6 caused by a slower interCT PIA decay, spectrally overlapping with the Y6 $T_1$) also showing dominant non-geminate recombination in these blends.

## 3. Transient Absorption Measurements on NFA Y7

**Halogenation of the Donor**

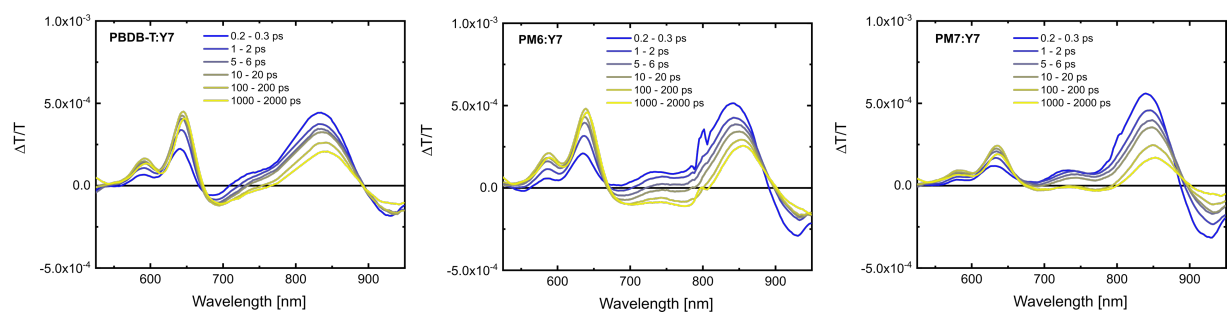

**Figure S4.** TA spectra for PBDB-T:Y7, PM6:Y7 and PM7:Y7. Excitation at 800 nm; fluence of ~0.6 µJ cm$^{-2}$.

Figure S3 shows TA measurements for PBDB-T:Y7, PM6:Y7, and PM7:Y7. All samples show comparable Y7 GSB at 840 nm, indicating a comparable number of excitations generated on Y7. While the polymer GSB in PBDB-T:Y7 and PM6:Y7 is visible after 0.2 ps, the polymer GSB in PM7:Y7 is very weak. Comparing the intensities of the polymer GSB when the hole transfer is completed at about 100 ps, the hole transfer yield in PBDB-T:Y7 and PM6:Y7 is comparable, with PM7:Y7 exhibiting the lowest yield. PM7:Y7 also reveals the fastest decay of Y7 GSB, indicating the highest amount of undissociated $S_1$ excitons decaying to the ground state. Overall, the hole transfer yield of Y7 blends is lower than that of Y6 blends, which may be due to the fact that CF is not the optimal solvent, as will be discussed in the next section. Comparing the hole transfer kinetics of the polymer GSB between 580 and 650 nm of all blends with NFA Y7 together with the kinetics of Y6 blends (Figure S5), the hole transfer kinetics for all six blends are comparable.

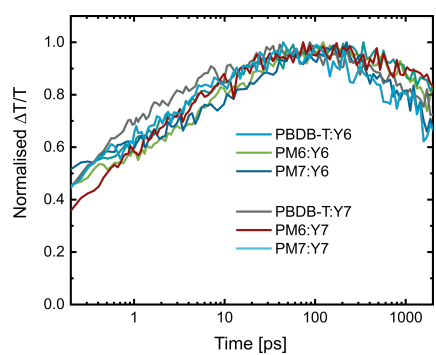

**Figure S5.** Hole transfer kinetics for all PM:Y series blends. Hole transfer rates are extracted from the polymer GSB 580 – 650 nm. All blends show comparable hole transfer kinetics.

**Influence of the Solvent**

The stacking and crystallinity of the NFAs Y6 and Y7 strongly depends on the solvent. Y6 shows a more dominant crystalline orientation and a better π-π stacking in CF than in chlorobenzene (CB).[18] However, the chlorination in NFA Y7 gives the molecule a stronger tendency to aggregate. Thereafter a solvent with a higher boiling point, such as CB is suitable.[19] The optimized recipe for PM6:Y7 solar cells is thus rather obtained with CB than CF as solvent.[11, 19]

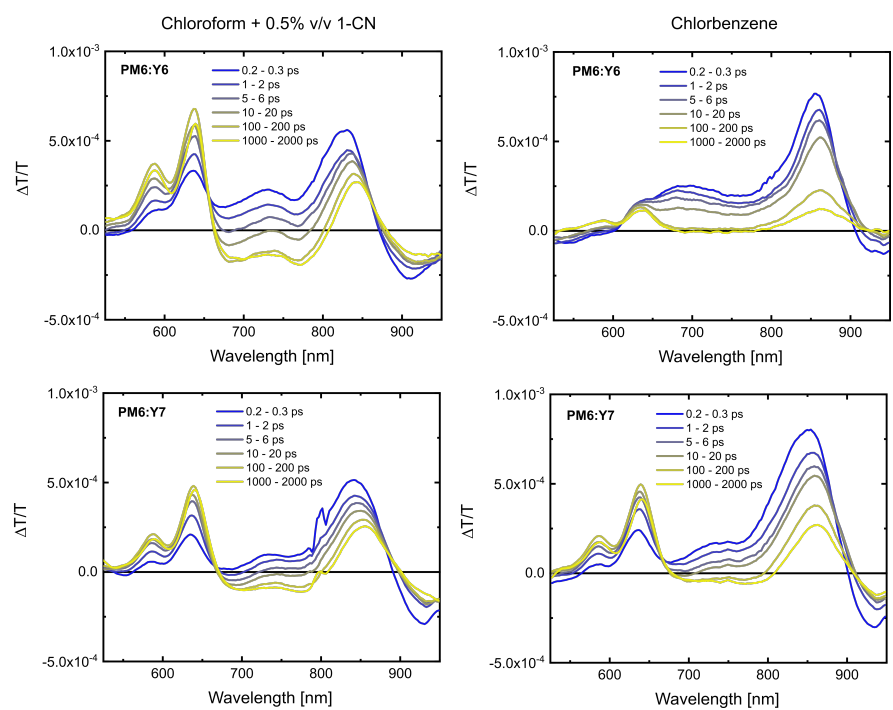

**Figure S6.** TA measurements for PM6:Y6 (top) and PM6:Y7 (bottom) dissolved in CF/1-CN (left) and CB (right). While the hole transfer yield decreases in PM6:Y6 when exchanging CF with CB, the hole transfer yield is comparable in PM6:Y7. Excitation at 800 nm; fluence of ~0.6 µJ cm$^{-2}$.

Figure S5 shows TA measurements for PM6:Y6 (top) and PM6:Y7 (bottom) dissolved in CF/ 1-CN (left) and CB (right). While PM6:Y6 in CF shows a PM6 GSB already at 0.2 ps, significantly increasing up to 100 ps, PM6:Y6 in CB shows a considerable smaller PM6 GSB, first arising after 10 ps and not increasing in time. While the Y6 GSB is initially comparable, it decreases quickly in PM6:Y6 dissolved in CB, indicating a high amount of undissociated excitons. While the hole transfer yield in PM6:Y6 significantly decreases when going from CF to CB, the hole transfer yield in PM6:Y7 is approximately comparable. These results agree with a reduced short-circuit current density $j_{SC}$ in PM6:Y6 and enhanced $j_{SC}$ in PM6:Y7 upon fabrication with CB instead of CF.[18, 19] Thus, when using the optimized recipe for Y7 blends (in Figure S5 without 1-CN), Y7 blends could turn out to be better with this fabrication method as shown for the solar cells.

## 4. Computational Details

The gas-phase ground state equilibrium geometry of three donor tetramers (PBDB-T, PM6 and PM7) and three NFAs (Y5, Y6, Y7) was optimized at the DFT level, using the exchange-correlation functional ωB97X-D and the basis set 6-31G(d,p). In order to speed up the calculations, the alkyl chains were replaced with methyl groups, both in donors and NFA Y-series.[20] Then, singlet and triplet vertical excitation energies were assessed by means of time-dependent DFT calculations within the Tamm-Dancoff approximation (TDA).[21] For such TDA TDDFT calculations, we resorted to a *screened* range-separated hybrid (SRSH) LC-ωhPBE functional and the 6-311G(d,p) basis set, as previously done elsewhere.[14] In the SRSH approach, solid-state screening effects are introduced by adjusting two parameters, α and β in addition to the non-empirical tuning of the exchange range-separation parameter ω.[22] The optimal gas-phase ω value was found following the gap-tuning procedure.[23] Then, the dielectric constant of the surrounding medium was chosen at ε = 4.5, $\alpha$ was fixed at 0.2, and, according to the relationship $\alpha + \beta = \frac{1}{\varepsilon}$, $\beta$ was set at 0.0222. All quantum-chemical calculations were carried out with the Gaussian16 suite.[24]

| Material | $f$ (osc. str.) | $S_1$ [eV] | $T_1$, $T_2$ [eV] | $\Delta E_{ST}$ [eV] |
|---|---|---|---|---|
| PBDB-T | 5.031 | 2.21 | 1.81 | 0.40 |
| PM6 | 5.304 | 2.24 | 1.83 | 0.41 |
| PM7 | 5.398 | 2.24 | 1.83 | 0.41 |
| Y5 | 2.481 | 2.02 | 1.46 | 0.56 |
| Y6 | 2.461 | 2.00 | 1.44, 1.65 | 0.56, 0.35 |
| Y7 | 2.505 | 1.97 | 1.42, 1.63 | 0.55, 0.34 |

**Table S2.** Calculated singlet and triplet vertical excitation energies, including the oscillator strength $f$ for singlet transitions. The computed $\Delta E_{ST}$ gap was then subtracted from experimental singlet energy values to accurately estimate $T_1$ and $T_2$ energies, as reported in Table S1.

The parameters that enter the Marcus-Levich-Jortner equation (see main text) were computed as follows. As regards the internal reorganization energy $\lambda_i$, we used the 4-points method, having performed gas-phase excited-state optimizations of $S_1$ and $T_1$ of the NFA Y-series at TDA TDDFT level, by employing the ωB97X-D functional and the 6-31G(d,p) basis set. Then, the energy differences $\Delta E^0_{S1-Tx}$ (where $x$ = 1 and 2) were evaluated at the SRSH TDA TDDFT LC-ωhPBE/6-311G(d,p) level of theory on the excited-state optimized structures (eventually, we also optimized the $T_2$ state for this purpose). SOC matrix elements were computed in the Brett-Pauli spin-orbit Hamiltonian framework as implemented in the PySOC code.[25] These calculations, on the other hand, were performed with the ωB97X-D functional and expanding the basis set to a Def2-TZVP, taking as input geometry the NFA Y-series $S_1$ optimized structure.

| NFA | $\Delta E^0_{S1-T1}$ [eV] | $<S_1|H_{SOC}|T_1>$ [eV] | $\Delta E^0_{S1-T2}$ [eV] | $<S_1|H_{SOC}|T_2>$ [eV] | $\lambda_i$ [eV] |
|---|---|---|---|---|---|
| Y5 | 0.59 | 7x10$^{-6}$ | 0.36 | 1x10$^{-5}$ | 0.040 |
| Y6 | 0.58 | 6x10$^{-6}$ | 0.35 | 1x10$^{-5}$ | 0.034 |
| Y7 | 0.57 | 6x10$^{-6}$ | 0.34 | 1x10$^{-5}$ | 0.032 |

**Table S3.** Calculated adiabatic energy differences between S$_1$ and the first two low-lying triplet states (T$_1$ and T$_2$), along with their corresponding SOC: $V_{SOC} = <S_1|H_{SOC}|T_x>$. The internal reorganization energy $\lambda_i$ is also reported.

| $\lambda_s$ [eV] | 0.05 | 0.10 | 0.15 | 0.20 | |
|---|---|---|---|---|---|
| Material | $\kappa_{ISC}$ [s$^{-1}$] | | | | |
| Y5 | 2x10$^5$ | 5x10$^5$ | 1x10$^6$ | 2x10$^6$ | 7x10$^4$ |
| Y6 | 2x10$^5$ | 5x10$^5$ | 1x10$^6$ | 2x10$^6$ | 4x10$^4$ |
| Y7 | 2x10$^5$ | 5x10$^5$ | 1x10$^6$ | 2x10$^6$ | 4x10$^4$ |

**Table S4.** Computed ISC rates from S$_1$ to T$_2$ as a function of the external reorganization energy $\lambda_s$. For the sake of completeness, we also report in red the ISC rates from S$_1$ to T$_1$, computed with $\lambda_s = 0.2\ eV$.

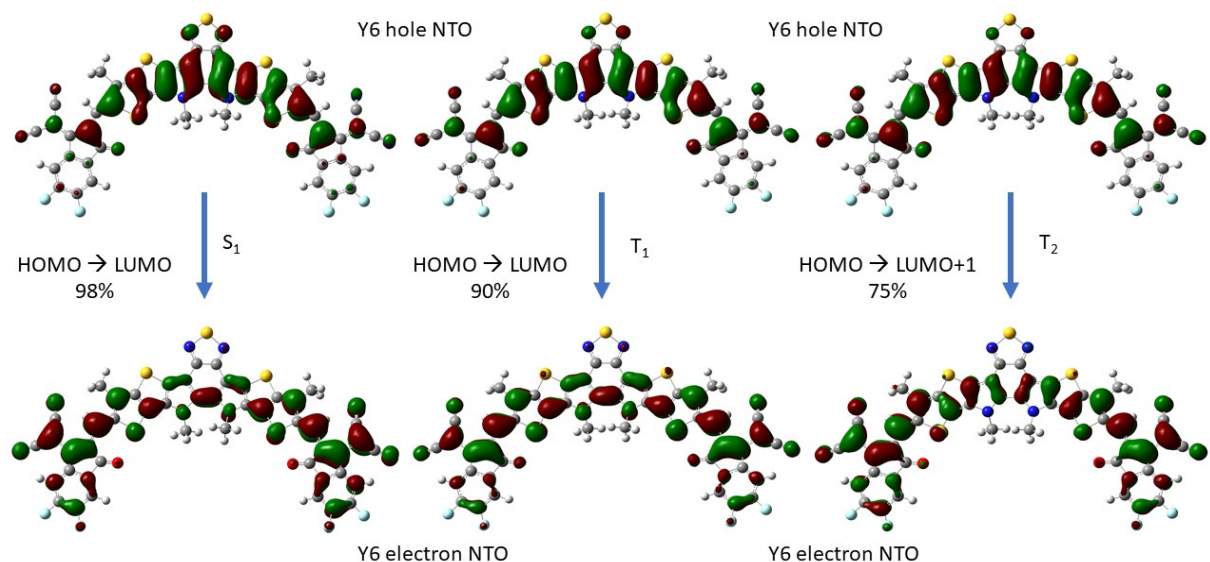

**Figure S7.** Hole and electron natural transition orbitals (NTOs) of Y6 as obtained at the SRSH TDA TDDFT LC-ωhPBE/6-311G(d,p) level of theory. The same NTOs were found for Y7, meaning that halogenation does not alter the nature of the low-lying excited states, namely S$_1$ and T$_2$ from which ISC can occur.

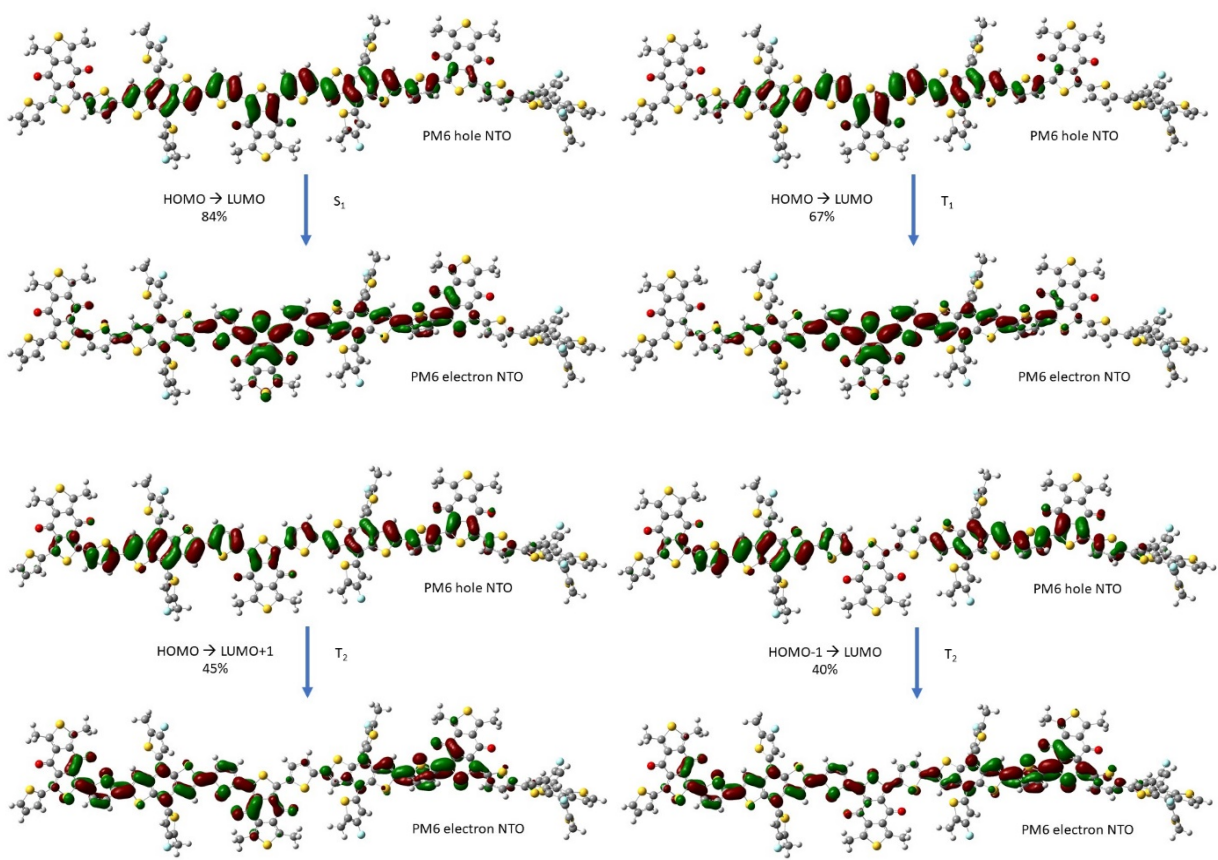

**Figure S8.** Hole and electron natural transition orbitals (NTOs) of a PM6 polymer chain as obtained at the SRSH TDA TDDFT LC-ωhPBE/6-311G(d,p) level of theory.

## 5. Additional PLDMR Data

The following part shows all PLDMR spectra of all polymers on spin-coated substrates (Figure S8) and PLDMR of dropcast samples of Y6 (Figure S10) and Y7 (Figure S11). The same presence of NFA triplet excitons can be found, whereby the width of the spectra are almost identical for dropcast and spin-coated samples. Only the ordering factor is higher for spin-coated substrates, as the molecules show preferential orientation on the substrate (Table S5 and S6). NFAs thereby stack face-on on the substrate, while the donors show a coexistence of edge-on and face-on orientation, which is visible by the shoulder in the HF signal and the additional smaller wings at 315 and 360 mT (Figure S8).[18, 26, 27] In the blends, the face-on stacking of NFA is maintained.[28]

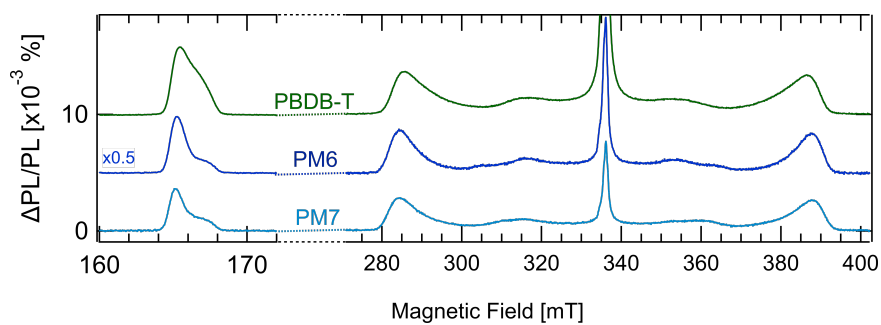

**Figure S9.** PLDMR spectra of the neat PBDB-T, PM6, PM7 on spin-coated substrates. The position of the HF signal and the width of the FF signal are similar (ZFS parameter in Table S5). Additionally, all PLDMR spectra show a considerable ordering (Table S5).

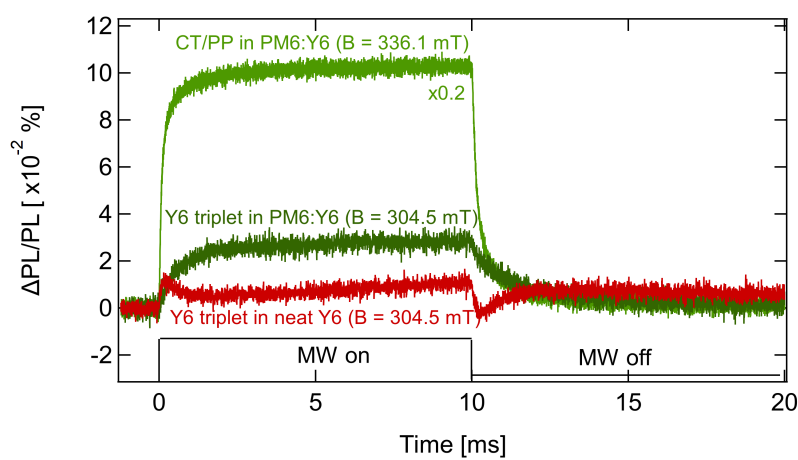

**Figure S10.** Transient PLDMR (trPLDMR) at distinct magnetic field values for neat Y6 (red) and PM6:Y6 (green) on spin-coated substrates. The transient response of the Y6 triplet signal in blended film PM6:Y6 (dark green, $B$ = 304.5 mT) is similar to the CT/PP feature in PM6:Y6 (light green, $B$ = 336.1 mT), while neat Y6 triplet shows a different temporal response to magnetic resonance conditions (red, $B$ = 304.5 mT). The temporal PLDMR response supports that Y6 triplets in PM6:Y6 are occupied via CT intermediate steps after non-geminate recombination, while neat Y6 triplets are predominantly populated via ISC. All signals show positive sign of the PLDMR signal.

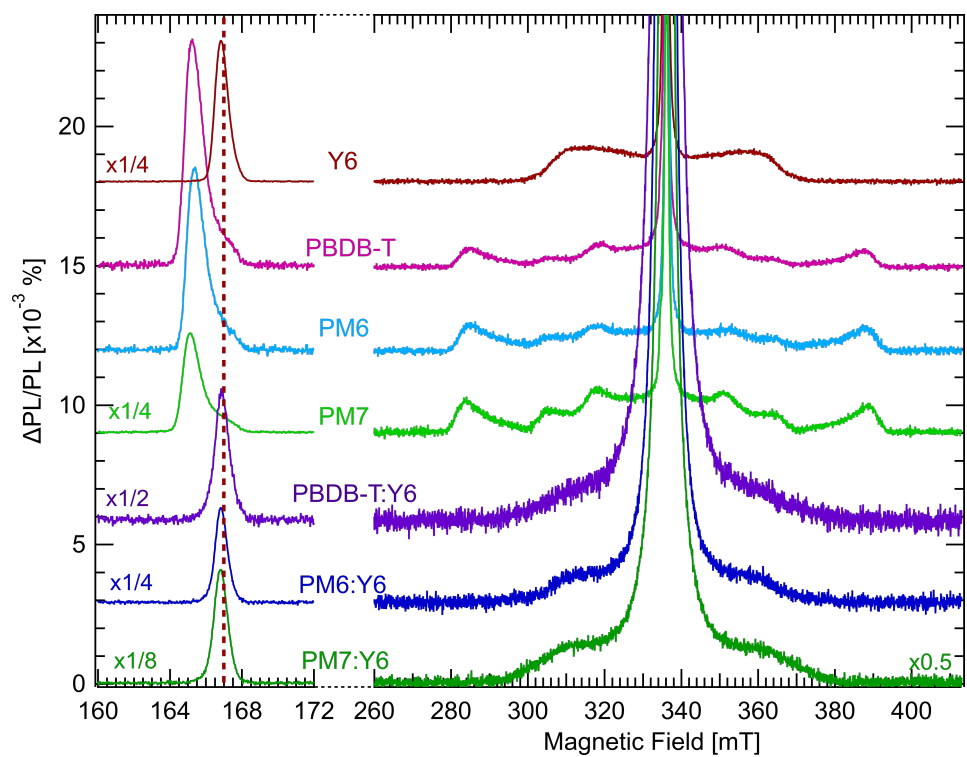

**Figure S11.** PLDMR spectra of the neat materials Y6, PBDB-T, PM6, PM7 and their blends PBDB-T:Y6, PM6:Y6 and PM7:Y6. The position of the HF signal and the width of the FF signal reveal Y6 triplet excitons in all blends.

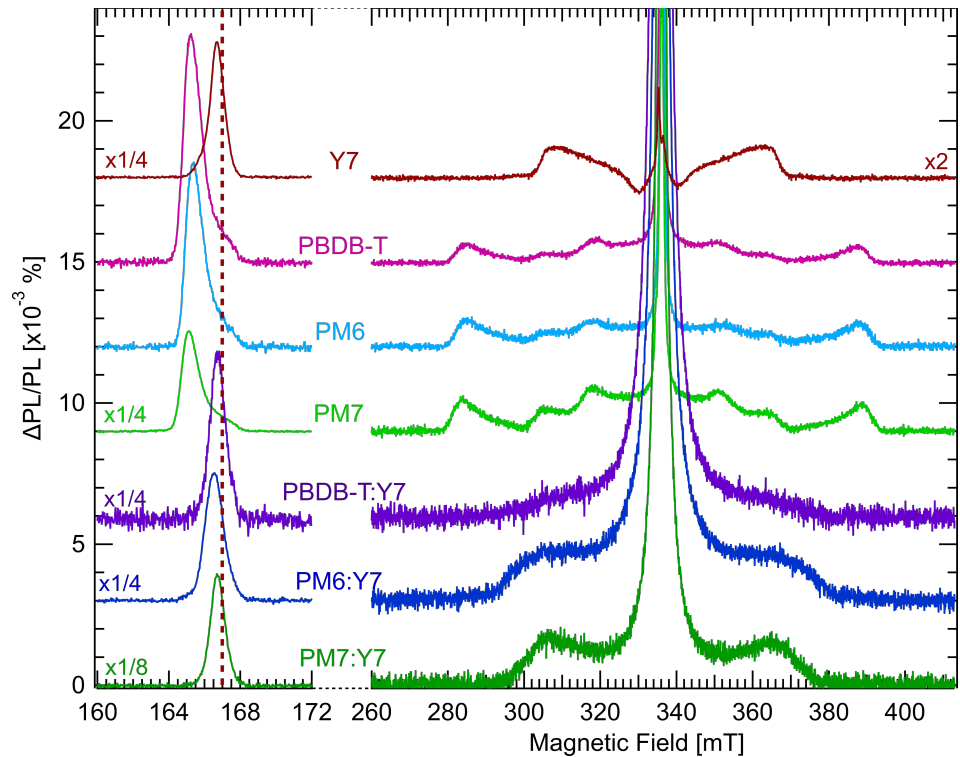

**Figure S12.** PLDMR spectra of neat materials Y7, PBDB-T, PM6, PM7 and their blends PBDB-T:Y7, PM6:Y7 and PM7:Y7. The position of the HF signal and the width of the FF signal reveal Y7 triplet excitons in all blends.

## 6. trEPR Signal Pattern and Additional Data

**trEPR Pattern by SOC-Driven ISC**

Spin-orbit coupling (SOC) is the dominant mechanism for ISC in small molecules since it decays rapidly with electron-hole distance with typical coupling values of few meV. At larger electron-hole distances, ISC is driven by the electron-hole distance-independent hyperfine interaction (HFI) which is typically negligible compared to SOC in small molecules.[29] The trEPR pattern arising is drawn in Figure S3 for Y6 with zero-field populations [$p_z$, $p_y$, $p_x$] = [0, 0.66, 0.34]. SOC-driven ISC acts on the zero-field triplet states, given by the eigenstates of the Hamiltonian |X⟩, |Y⟩ and |Z⟩, whose populations can be transferred to the populations of the high-field states |T$_+$⟩, |T$_0$⟩, |T$_-$⟩.[30, 31] The splitting of the triplet sublevels thereby depends on the orientation of the external magnetic field $\vec{B}$ with respect to the principal axes X, Y, and Z of the ZFS tensor.[30] The energy of the aligned state, called canonical orientation, is thus independent of $B$ and equal to the energy at zero field, while the energies of the other two states increase/decrease with magnetic field (Figure S12).[31] The population of the high-field state is therefore different for each canonical orientation:[30, 31]

$$\vec{B}||Z: p_0 = p_z, p_\pm = 0.5\,(p_x + p_y)$$

$$\vec{B}||Y: p_0 = p_y, p_\pm = 0.5\,(p_x + p_z)$$

$$\vec{B}||X: p_0 = p_x, p_\pm = 0.5\,(p_y + p_z)$$

For each orientation of the principal axis of **D** with respect to $\vec{B}$, there are enhanced absorptive and emissive transitions in magnetic resonant conditions (Figure S12), which add up in rigid samples to a characteristic powder pattern.[30]

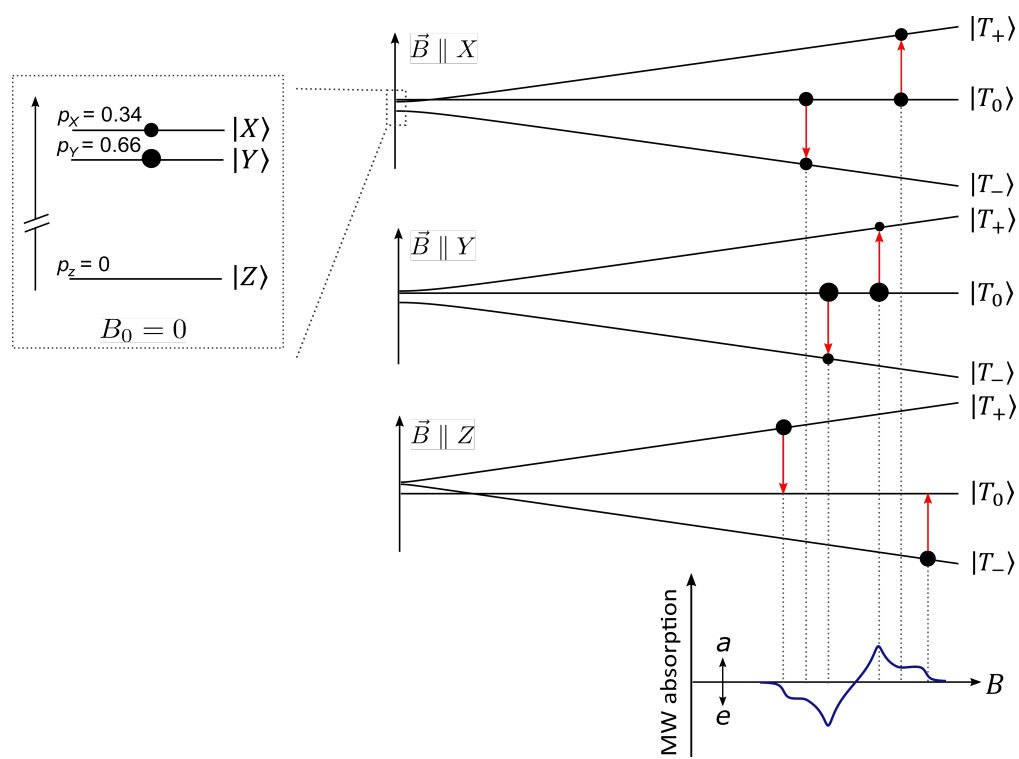

**Figure S13.** Generation of a Y6 trEPR spectrum with [$p_z$, $p_y$, $p_x$] = [0, 0.66, 0.34]. ISC is sublevel-selective and acts on the zero-field sublevels |X⟩, |Y⟩ and |Z⟩. These populations are converted into high-field populations |T$_+$⟩, |T$_0$⟩ and |T$_-$⟩, depending on the principal axes (X, Y, Z) orientation of the ZFS tensor **D** with respect to magnetic field $\vec{B}$. The aligned state is thereby energetically independent of the external magnetic field, while the other two states increase/decrease with *B*. The unequal populations result in microwave emission (*e*) or absorption (*a*) for Δ$m_s$ = ±1 transitions (red arrows) for each canonical orientation. The spectra of all orientations of **D** relative to $\vec{B}$ add up to a characteristic trEPR signal (blue). (adapted from Ref.[30])

**trEPR Pattern by Geminate Back Charge Transfer**

Molecular triplet excitons populated by geminate back charge transfer (BCT) can be distinguished from SOC-driven ISC by another pattern in the trEPR signal. The spin polarization on the triplet sublevels is thereby converted from the CT states. As already explained in the main text, CT states can be explained in the context of a spin-correlated radical pair (SCRP) whereby spin mixing occurs only between states $|^3CT_0\rangle$ and $|^1CT_0\rangle$ ($m_s$ = 0). Thus, regarding the triplet sublevels, only $|^3CT_0\rangle$ is occupied. If charge transfer excitons undergo spin-allowed BCT to triplet excitons with shorter spin-spin distances, i.e., molecular triplet states, the spin polarization is preserved. The overpopulated |T$_0$⟩ state thus leads to an *aeeaae* or *eaaeea* pattern as shown in Figure S13. This pattern can only be obtained by selective population of high-field spin states and can thus be distinguished from triplet excitons occupied by SOC-driven ISC.

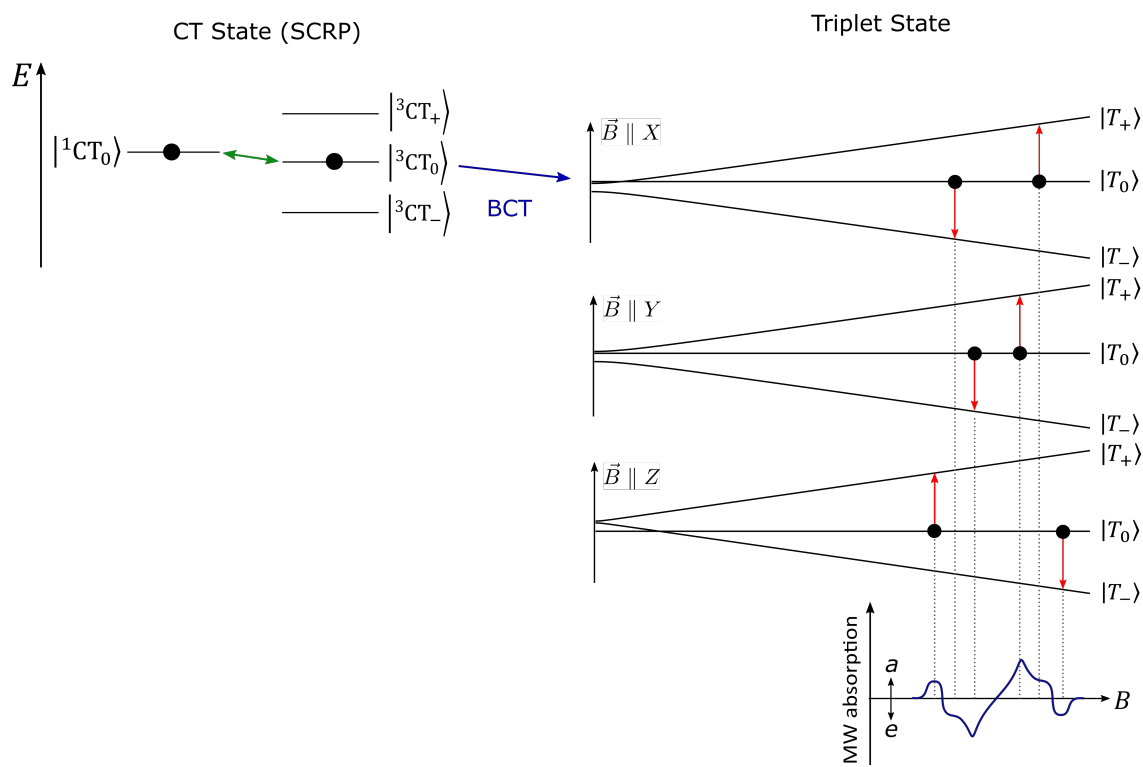

**Figure S14.** Generation of a trEPR spectrum generated by geminate BCT. Due to angular momentum conservation, spin mixing occurs only between $|^3CT_0\rangle$ and $|^1CT_0\rangle$ ($m_s = 0$). Spin-allowed BCT to triplet excitons preserves spin polarization, whereby high-field $|T_0\rangle$ states get overpopulated, arising here in an *aeeaae* pattern.

**Additional trEPR Spectra**

The following part shows trEPR spectra of polymers (Figure S14) and Y7 including blends (Figure S15).

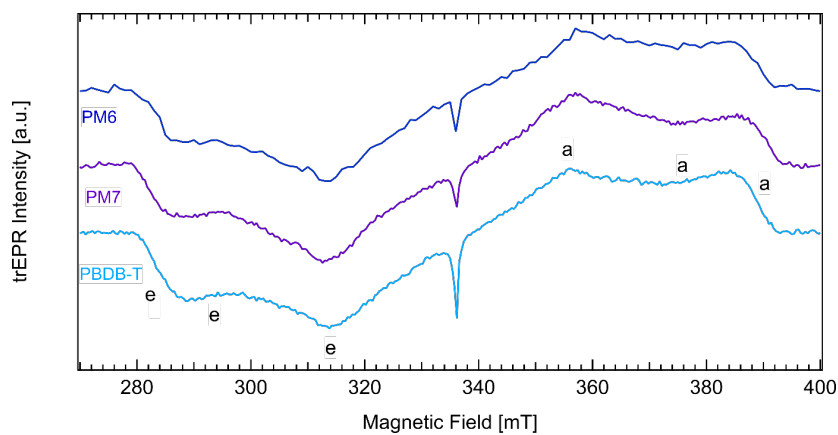

**Figure S15.** trEPR of neat PM6, PM7 and PBDB-T. All spectra show a SOC-driven ISC pattern (*eeeaaa*) and a small CT contribution at 336.1 mT. Simulations including parameters are given below.

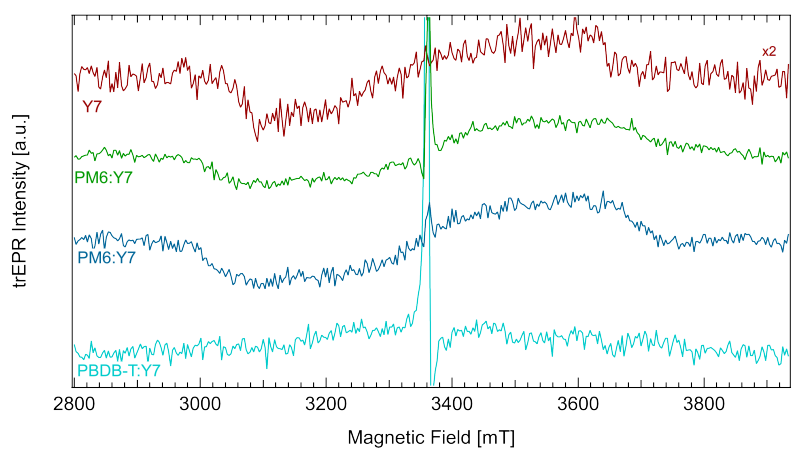

**Figure S16.** trEPR of neat Y7 and the blends PM6:Y7, PM7:Y7 and PBDB-T:Y7. All triplet spectra show a SOC-driven ISC pattern (*eeeaaa*). PBDB-T:Y7 shows negligible triplet yield by ISC, which is comparable to PBDB-T:Y6.

## 7. HOMO/LUMO Level

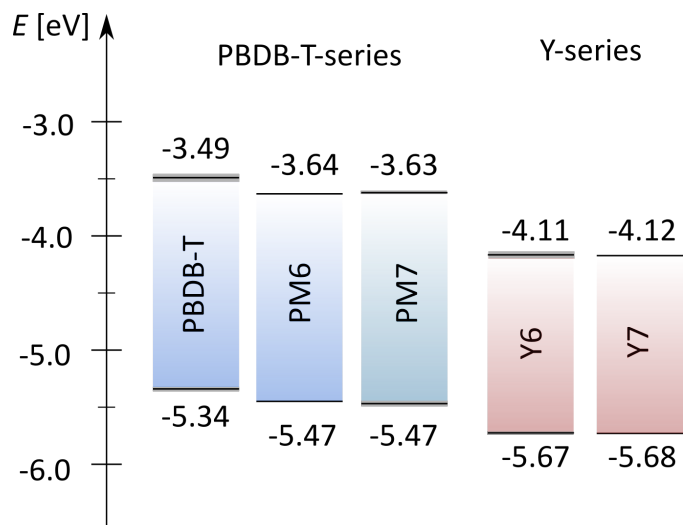

**Figure S17.** HOMO and LUMO levels of donors PBDB-T, PM6 and PM7 as well as acceptors Y6 and Y7. The HOMO of halogenated donors PM6 and PM7 are downshifted in comparison to non-halogenated PBDB-T. Thus, the HOMO offset and therefore the driving force for hole transfer of halogenated OPV blends PM6:Y6, PM7:Y6, PM7:Y6 and PM7:Y7 is lower than for PBDB-T:Y6 and PBDB-T:Y7. The values are given with the mean value and standard error taken from literature for PBDB-T[32-35], PM6[32, 34-37], PM7[36-39], Y6[13, 40-43], Y7[44, 45].

## 8. Triplet Simulation Parameters

**PLDMR**

|          | Triplet |  |  |  |  | CT |  |
|----------|---------|---|---|---|---|---|---|
| Material | D/h [MHz] | E/h [MHz] | λ_θ, λ_φ | Lw [mT] | weight | Lw [mT] | weight |
| Y6 | 950 | 150*<br>150 | 11.0, 0<br>0.0, 0 | 4.0, 0<br>2.0, 0 | 0.62<br>0.11 | 1, 2.1<br>0, 1.3 | 0.27<br>-0.06 |
| PBDB-T | 1500 | 120<br>120 | 7.1, 0<br>-1.5, 4.0 | [4.5 0]<br>[0 8.9] | 0.37<br>0.27 | [1 1.4] | 0.36 |
| PM6 | 1510 | 120<br>120<br>75 | 11.4, 0<br>-2.0, 8.0<br>-1.0, -5.0 | [4 0]<br>[0 8]<br>[0 5] | 0.40<br>0.12<br>0.28 | [0 1] | 0.20 |
| PM7 | 1510 | 120<br>120<br>50 | 8.0, 0<br>-2.0, 7.0<br>-2.0, 0 | [4 0]<br>[0 8]<br>[0 7] | 0.48<br>0.20<br>0.20 | [0 1.4] | 0.12 |
| PBDB-T:Y6 | 1040 | 200* | 4.5, 0 | [7 0] | 0.20 | [0.8 1.9] | 0.80 |
| PM6:Y6 | 1020 | 200* | 5.5, 0 | [7 0] | 0.15 | [0.8 2.2] | 0.85 |
| PM7:Y6 | 1000 | 200* | 5.5, 0 | [7 0] | 0.38 | [0.9 2.5] | 0.62 |

**Table S5.** Parameter for PLDMR spectral simulations using the MATLAB tool EasySpin for spin-coated substrates. *: $E$ value cannot be determined due to high ordering.

|          | Triplet |  |  |  |  | CT |  |
|----------|---------|---|---|---|---|---|---|
| Material | D/h [MHz] | E/h [MHz] | λ_θ, λ_φ | Lw [mT] | weight | Lw [mT] | weight |
| Y6 | 830 | 150 | 2.0, -1.0 | [8 0] | 0.56 | [1.1 0.8]<br>[0 0.8] | 0.44<br>-0.19 |
| Y7 | 1040 | 230 | 0.5, -0.7 | [0 1.5] | 0.97 | [0 1.2]<br>[0 0.83] | 0.03<br>-0.02 |
| PBDB-T | 1500 | 185<br>185<br>70 | 9.0, 0.0<br>-0.4, 2.8<br>-2.2, -1.0 | [3.5 0]<br>[3.5 0]<br>[3 0] | 0.13<br>0.3<br>0.07 | [0 1.5] | 0.51 |
| PM6 | 1510 | 170<br>170<br>75 | 8.0, 0.0<br>0.0, 3.5<br>-1.0, -2.5 | [4 0]<br>[0 3.5]<br>[0 3.5] | 0.1<br>0.66<br>0.02 | [0.3 0.9] | 0.22 |
| PM7 | 1530 | 200<br>200<br>80 | 9.0, 2.0<br>-0.6, 2.3<br>-4.5, -1.7 | [4 0]<br>[4 0]<br>[3.5 0] | 0.14<br>0.42<br>0.08 | [0 1.5] | 0.36 |
| PBDB-T:Y6 | 920 | 230 | 0.6, 2.2 | [10 0] | 0.03 | [1.2 0.9] | 0.97 |
| PM6:Y6 | 850 | 240 | 0.6, 0.2 | [5 0] | 0.04 | [0.8 1.1] | 0.96 |
| PM7:Y6 | 900 | 220 | 1.6, 2.0 | [15 0] | 0.16 | [0.6 1.9] | 0.84 |
| PBDB-T:Y7 | 1060 | 205 | 0.6, 0.6 | [9 0] | 0.06 | [0.5 1.5] | 0.94 |
| PM6:Y7 | 1100 | 240 | 1.2, 1.3 | [9 0] | 0.26 | [0.7 1.8] | 0.74 |
| PM7:Y7 | 1190 | 240 | -1.0, -3,0 | [9.5 0] | 0.26 | [1.0 2.0] | 0.74 |

**Table S6.** Parameter for PLDMR spectral simulations using the MATLAB tool EasySpin for dropcast samples.

**Table S5 and S6** list the EasySpin simulation parameters for PLDMR spectra of triplet excitons and CT states in the various pristine materials and blends. Parameters given are the ZFS parameters $D$ and $E$, ordering factors $\lambda_\theta$, $\lambda_\varphi$, the spectral linewidth with Gaussian and Lorentzian contributions [Gaussian Lorentzian] and the weight of the signals (relative signal intensity). The ordering factor $\lambda$ specifies the orientational distribution of the molecules in the sample and is reflected by "wings" in the PLDMR spectrum.[46, 47] The ordering is more pronounced in PLDMR than in trEPR, as seen in previous studies.[48] The ordering factor $\lambda$ is given for θ and φ, whereby θ is the angle between the molecular z-axis and static magnetic field $B_0$ and φ is the in-plane angle. If the ordering factor $\lambda$ is zero, all molecular orientations occur with the same probability. The weight is the relative intensity of the spectral contributions of superimposed triplet and CT spectra.

The polymers PBDB-T, PM6 and PM7 show triplet excitons with different ordering: All triplet excitons possess the same dipolar interaction ($D$ value), whereby we found two different $E$ values, i.e., value for the rhombicity or the non-axial symmetry, indicating different internal arrangement.[49] These different triplet excitons have a preferred orientational distribution, given by the ordering factors. The NFAs possess a broader CT peak (comparable to blends) and a small negative CT contribution. All blends possess a high weight in the CT peak due to additional CT states. All the blends with NFA Y7 show triplet contribution on the NFA. The $D$ value of Y7 and its blends is higher than NFA Y6 and its blends.

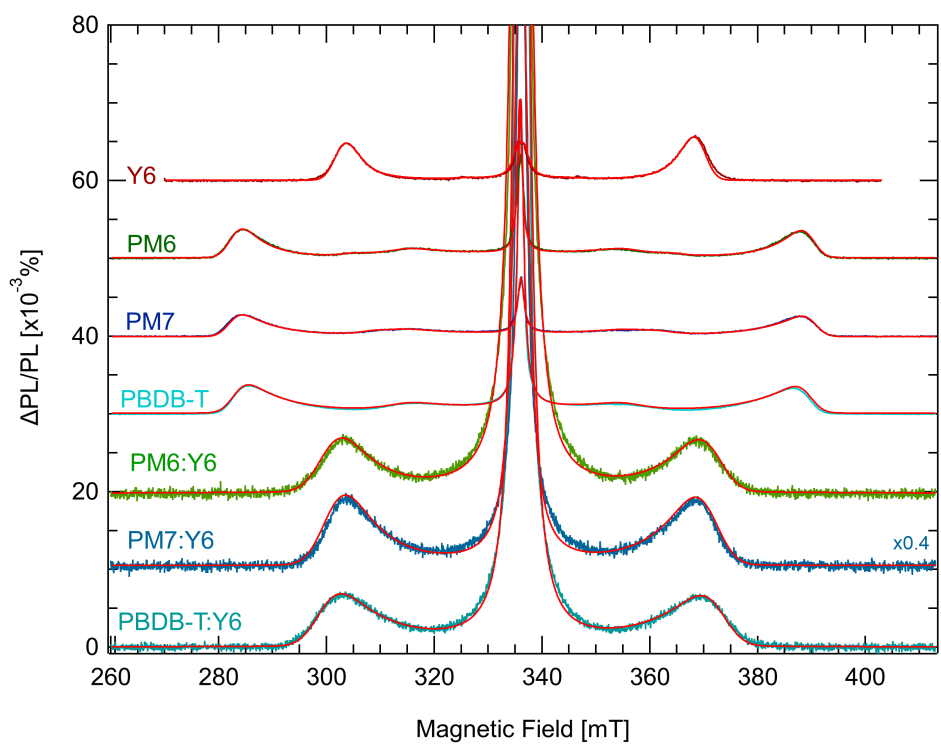

**Figure S18.** EasySpin simulations (red) of all PLDMR measurements on spin-coated substrates (colored).

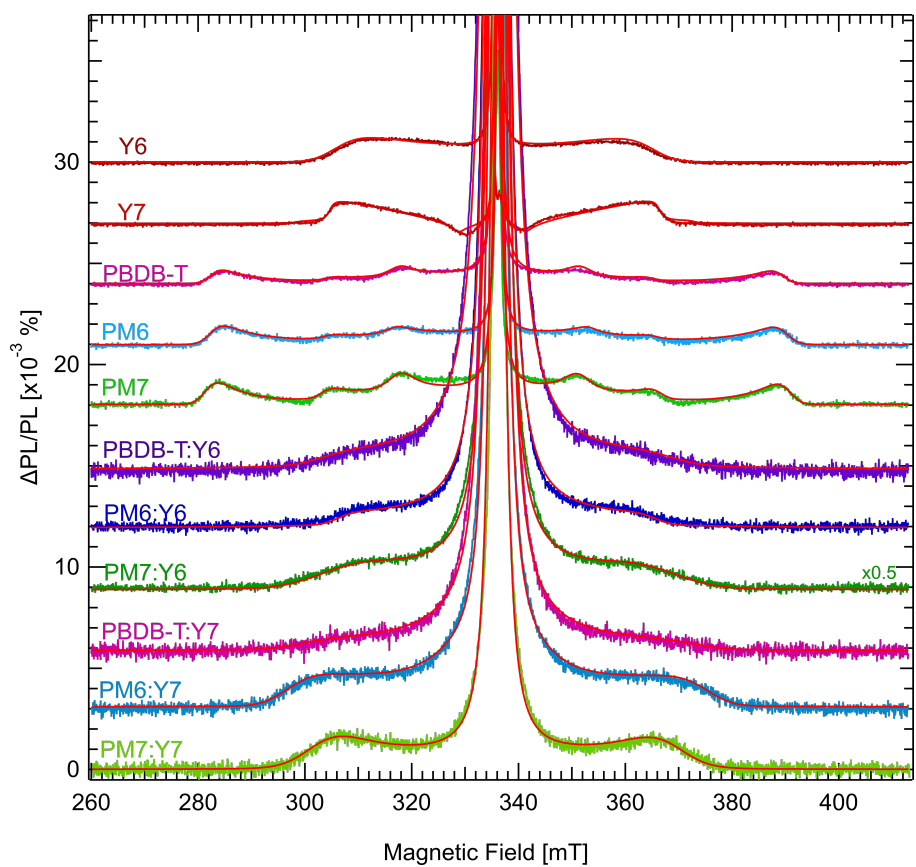

**Figure S19.** EasySpin simulations (red) of all PLDMR measurements (colored).

**trEPR**

| Material | $D$/h [MHz] | $E$/h [MHz] | $[p_z, p_y, p_x]$ | $\lambda_\theta, \lambda_\varphi$ | $Lw$ [mT] |
|---|---|---|---|---|---|
| Y6 | 945 | 215 | [0.00 0.66 0.34] | 0, 0 | [0 2] |
| Y7 | 850 | 240 | [0.00 0.62 0.42] | 0, 0 | [0 1.5] |
| PBDB-T | 1460 | 95 | [0.00 0.51 0.49] | 0.68, 0.44 | [0 2.3] |
| PM6 | 1455 | 100 | [0.00 0.56 0.44] | 0.42, -0.1 | [0 2.7] |
| PM7 | 1510 | 120 | [0.00 0.53 0.47] | 0.35, 0.1 | [0 2.0] |
| PM6:Y6 | 995 | 210 | [0.00 0.66 0.34] | 0.42, -0.1 | [0 3.2] |
| PM7:Y6 | 880 | 160 | [0.00 0.66 0.34] | 0.35, 0.1 | [0 2.0] |
| PM6:Y7 | 900 | 200 | [0 0.60 0.40] | 0.42, -0.1 | [0 4] |
| PM7:Y7 | 925 | 200 | [0 0.60 0.40] | 0.35, -1.0 | [0 2.1] |

**Table S7.** Parameter for trEPR spectral simulations using the MATLAB tool EasySpin.

**Table S7** lists the EasySpin simulation parameters for trEPR spectra of triplet excitons in the various pristine materials and blends. Parameters given are the ZFS parameters $D$ and $E$, relative zero-field populations, ordering factors $\lambda_\theta$, $\lambda_\varphi$ and the spectral linewidth with Gaussian and Lorentzian contributions [Gaussian Lorentzian].

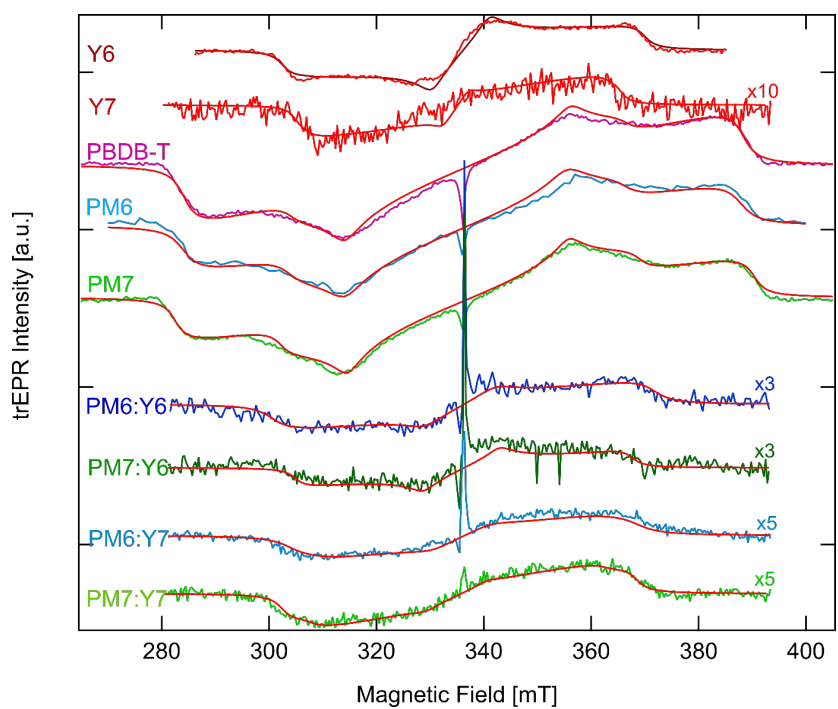

**Figure S20.** EasySpin simulations (red) of trEPR measurements (colored).